# The Magnetic Toroidal Sector as
# A broad-band Electron-Positron Pair Spectrometer
# I. Lepton Trajectories

S. Hagmann[1,2], P.M. Hillenbrand[1,3], Y. Litvinov[1], U. Spillmann[1], Th. Stöhlker[1,4,5]

[1]GSI Helmholtzzentrum Darmstadt, 64291 Darmstadt, [2]Inst. für Kernphysik, Universität Frankfurt, 60438 Frankfurt,

[3]Universität Giessen, 35392 Giessen, [4]Helmholtz Institut Jena, 07743 Jena, [5]Fakultät für Physik, Universität Jena, 07745 Jena

We report an analysis of electron-optical properties of a toroidal magnetic sector spectrometer and examine parameters for its implementation in a relativistic heavy-ion storage ring like HESR. For studies of free-free pair production in heavy- ion atom collisions this spectrometer exhibits very high efficiencies for coincident $e^+$-$e^-$ pair spectroscopy over a wide range of momenta of emitted lepton pairs. The high coincidence efficiency of the spectrometer is the key for stringent tests of theoretical predictions for the coincident positron- and electron emission characteristics and for the phase space correlation of lepton vector momenta in free-free pair production.

## 1. Introduction

Electron –positron pair production, the excitation of an electron with negative energy from the completely filled Dirac sea into an unoccupied state with either a discrete positive energy (resulting in bound-free pairs) or positive energy in a continuum (resulting in free-free pairs) has evolved into a central topic of QED in extreme fields as the coupling between the lepton field and the electromagnetic field is close to one [1-3]. Already in the early times of the development of quantum mechanics and QED this process was assessed in the context of the Dirac-, Breit-Wheeler- and vacuum-polarization-in-strong- electromagnetic-fields- fundamental processes, and also the highly interesting and very profound relation of this particular channel in fast ion-atom collisions with other inelastic processes, e.g. electron nucleus bremsstrahlung (eNBS), has been thoroughly investigated early on [4-9]. These seminal calculations on eNBS have provided then in turn guidance for particular aspects in the evaluation of transition probabilities in pair creation calculations [19]. It is within this context that pair production in ion-atom collisions with projectile energies ranging from the extreme adiabatic regime close to the Coulomb barrier ( i.e. below 6AMeV) up to energies in the extreme relativistic regime with collisions characterized by $\gamma \approx$ few 1000 has since been subject of numerous theoretical treatises and experimental studies [10-51]. Most recently the surprisingly high cross sections even exceeding 100kb observed for pair production in highly relativistic very heavy ion atom collisions have aroused new interest in this process even in the high-energy accelerator community as capture from pair production has been identified as a critical, potentially luminosity limiting process in relativistic heavy ion colliders like LHC and RHIC [13,19,21,55,56]. The enormous cross sections observed can be traced [1,3,13,19,21] to the large transverse electric fields $E_{transv} \sim \gamma$ (see fig. 1) which give rise to the unexpectedly high cross section $\sigma_{pair\ production}$ for free-free pair production.

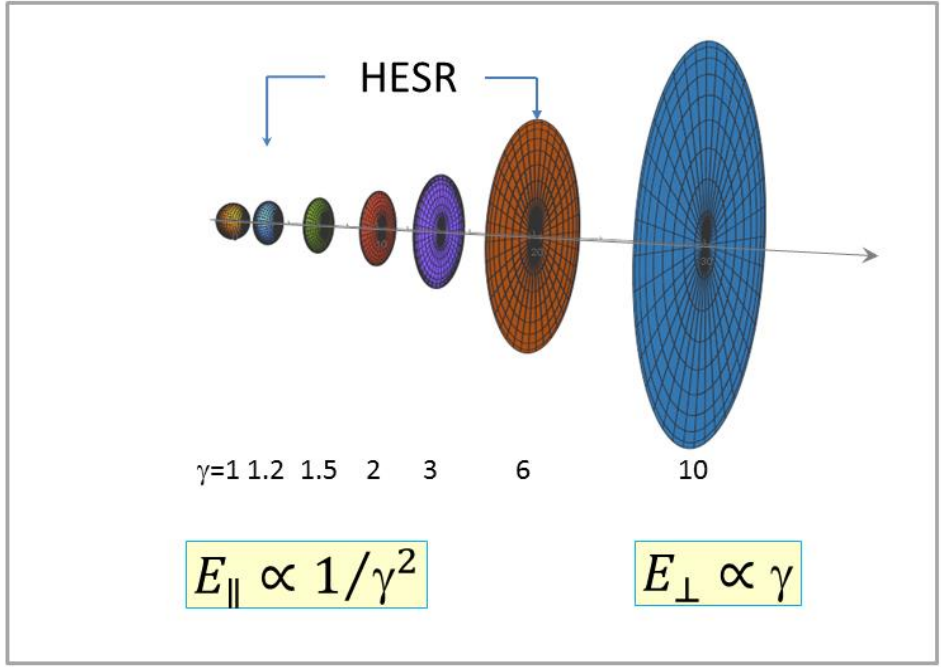

$\gamma = 1$ $1.2$ $1.5$ $2$ $3$ $6$ $10$

$$E_\parallel \propto 1/\gamma^2$$

$$E_\perp \propto \gamma$$

*Fig.1. Dominance of transverse electric fields for a moving charge with increasing $\gamma$. The current equivalent collision energy range of interest for the free-free pair production in the HESR is indicated. The strongly adiabatic region for positron production at $\approx 6$ AMeV with $\beta \approx 0.1$, i.e. $(\gamma-1) \ll 1$ and as well pair production in the extreme relativistic region with $\gamma \gg 100$ are not subject of our present considerations.*

The relativistic contraction of the collision time $\tau_{coll} = b/\gamma v_{coll}$ ( where b is the impact parameter) combined with the high transverse electric field $E_{trans} \sim \gamma Z_{proj}/b^2$ makes the electron positron pair-creation process in collisions of near relativistic heavy ions with atoms particularly attractive as a tool to investigate in high resolution correlated lepton dynamics for collision times $t_{coll} < 10^{-18}$s and at collision velocities $v_{coll}$ larger than or large in comparison with the orbital velocity $v_K$ of the inner most bound electron $v_{coll} > v_K$ .

In the entire discussion following this introduction we will only be concerned with collisions with $1.2 \leq \gamma \leq 6$, i.e. covering the range of collisions energies for ions stored in the future HESR storage ring [57-59]. We are not considering electron positron pair creation in those other two collision regions which also have attracted recent attention, due the new windows opened for fundamental atomic processes by intricate accel-decel modes in modern advanced accelerator facilities: these are very highly adiabatic collisions with $v_{coll} \ll v_K$ ( this corresponds to collision energies $\leq 6$AMeV, e.g for $U^{91+}$ projectiles), to be investigated at the future CRYRING/ESR, where during the collisions even a superheavy (i.e. $Z_1 + Z_2 = Z_{UA} \geq 173$) quasimolecule may be formed [1, 60-66], and second, ultraperipheral collisions with γ≈1000-3000, as now routinely available at CERN and RHIC [1, 14, 20, 21, 55].

At the future near relativistic storage ring HESR at FAIR the collision energy range up to γ≈6 corresponding to $E_{proj} \leq 5 AGeV$ will be available. For ion-induced pair-production in this collision velocity range theoretical calculations [5, 10, 11, 13-15, 18, 33, 36, 37, 41, 44, 46] have identified various major channels which may be distinguished experimentally (see fig .2),

i)      free-free pair production,

ii)     bound-free pair production and

iii)    negative continuum dielectronic recombination:

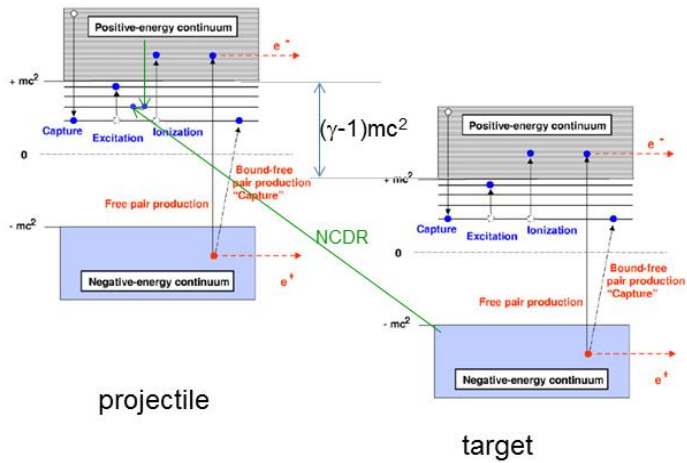

projectile

target

*Fig. 2. Energy diagram of the single particle Dirac equation and illustration of various open channels for electron-positron pair production besides the classical electron capture, ~excitation and ~ionization active in swift ion-atom collisions [57, 58, 63].*

The free-free pair creation described in eq.1a,1b is the main motivation for the current spectrometer design study strongly emphasizing a large combined detection efficiency (and solid angle) for simultaneous detection of electron and positron of a free-free pair: a coincident detection of both outgoing leptons and their vector momenta - as constitutive in the identification of free-free pair production - is not practically possible with conventional dispersive instruments covering only a small solid angle for each lepton because of the intrinsic very low pair coincidence efficiencies resulting from such configurations.

In free-free pair production both leptons are emitted into continuum states of $E_{kin} > 0$:

$$X^{Z+} + A \longrightarrow X^{Z+} + \{A^*\} + e^+ + e^- \qquad \textit{free-free} \qquad (1a)$$

Theory predicts a scaling of the total cross section for free-free pair production [10, 13, 36] following

$$\sigma_{free-free} \sim (Z_{proj}{}^2)(Z_{tar}{}^2)ln^3\gamma \quad . \qquad (1b)$$

For this channel a total cross section of ≈3b has been measured for 940 AMeV Au +Au collisions by Belkacem et al. [34, 40, 54]. The complexity of two-lepton continua here presents a formidable impediment to theoretical differential cross sections and as well a tremendous obstacle for experimental full differential cross sections (see next section).

For positron production from bound-free pair creation channels according to eqs. 2a and 2b, the differential cross sections on the other hand can be well studied using traditional magnetic spectrometers at the HESR storage ring [57-60, 80] as here only one lepton, the positron, is found in the continuum, with a resulting satisfactory overall coincidence efficiency when employing traditional magnetic spectrometers.

In the bound- free pair production channel the electron of the lepton pair is captured into a bound state of the projectile while the positron is emitted into the continuum.

$$X^{Z+} + A \longrightarrow X^{(Z-1)+}(nl) + \{A^*\} + e^+ \qquad (nl) \text{ mostly } (1s) \text{ ; bound-free} \qquad (2a)$$

Here theory predicts a scaling of the total cross section following

$$\sigma_{bound-free} \sim (Z_{proj}{}^5)(Z_{tar}{}^2)ln\gamma \qquad (2b)$$

and the total cross section, again as measured by Belkacem et al for 940 AMeV Au +Au collisions, is ≈2b [34, 40, 54].

The third important pair creation channel is the negative continuum dielectronic recombination (NCDR) channel as described in eq. 3 and has not been experimentally observed yet; it was only recently proposed [ 67-69]

$$X^{Z+} + A \longrightarrow X^{(Z-2)+} + \{A^{+*}\} + e^+ \qquad NCDR \qquad (3)$$

The NCDR channel has been predicted to exhibit a broad maximum as a function of $\gamma$ [60, 67-69] and not a monotonic dependence of the total collision cross section as established for free-free and bound-free pair production according to eqs. 1b and 2b. The NCDR is unambiguously characterized by a double capture into the projectile coincident with positron emission. Its total cross section maximizes at approximately 3mb at around 1AGeV U[92+] [67], this is much lower than the two other channels (1) and (2).

All three channels can be experimentally distinguished by their characteristics via $e^+/e^-$ $* X^{q+}$ coincidences, where the coincident projectile final charge state *is q= Z, Z-1, Z-2,* respectively.

## 2. Emission characteristics of free-free pair production in ion-atom collisions

The $e^+$-$e^-$-pair production in the field of a nucleus induced by a high energy photon is dominated by near back-to-back emission of both leptons and a near equal energy sharing $E_{e^+} \approx E_{e^-}$ of electron and positron [1]. For free-free pair production in the transient strong electro-magnetic field created in fast ion-atom collisions, however, the resultant emission patterns exhibit significantly more complex features (18,28,29).

In fig.3 we illustrate for the collision system 0.96AGeV $U^{92}$ + Au the total positron yield per $10^0$ interval as function of the laboratory observation angle as measured by Belkacem et al.[34]; the strong kinematic maximum at $1/\gamma$ rad is clearly visible.

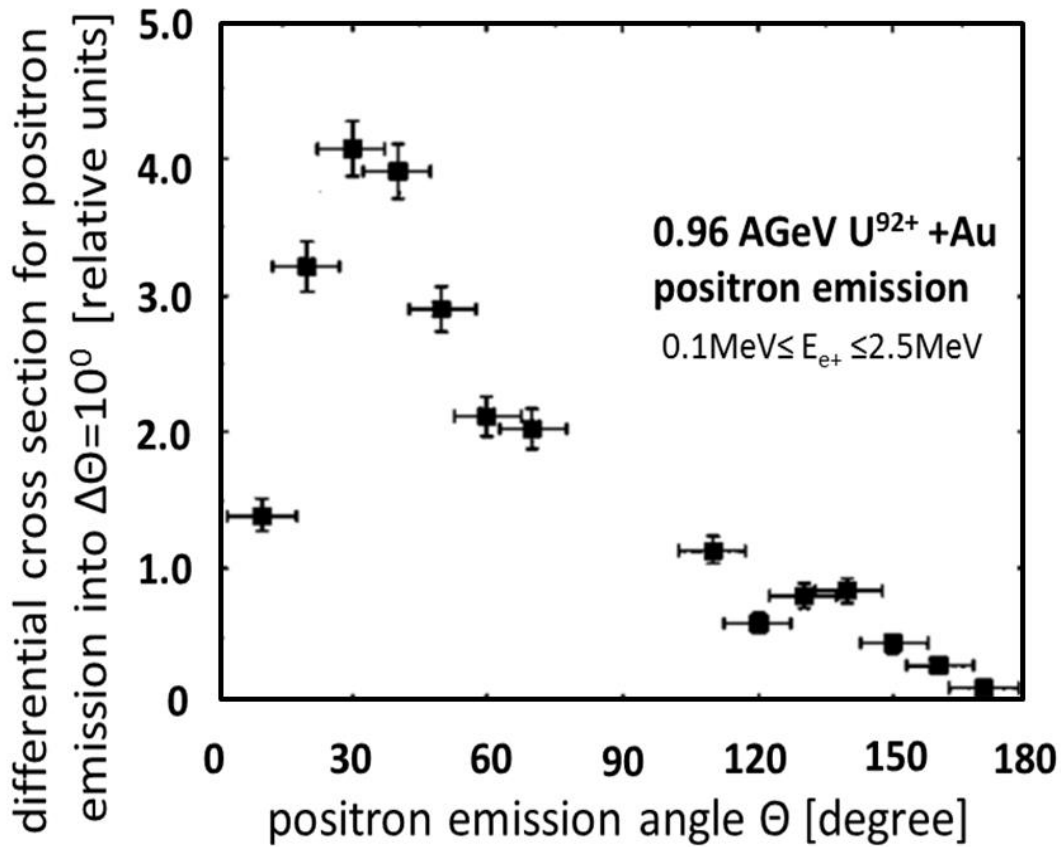

Fig. 3. Relative positron production cross section in $U^{92}$ + Au collisions at 0.96 AGeV, data are from A. Belkacem et al., ref. 34.

Few theoretical differential cross sections for free-free pair production have been reported [18, 28, 29, 36], however, and, as follows form the nature of the full 6-dimensional phase space and corresponding limitations for numerical computations, mostly for mutually not overlapping, and quite narrow subspaces. For the differential cross sections for pair creation in their fundamental relation to the high energy endpoint of enBS already Pratt et al. [50] and Fano et al. [51] in the early 60s have presented some general considerations pertaining to the single energy-differential cross sections with strongly different characteristics for high energy end of positron or electron spectra; these far reaching predictions have never been subject to experimental scrutiny.

For $e^+$-$e^-$-pair production in relativistic ion-atom collisions the group of W. Scheid [28, 29] performed extensive calculations for effectively double differential cross sections yielding some characteristics of free-free pair production. These authors used a semiclassical approach and solved the time dependent Dirac equations in the semiclassical approach employing free wavepackets in a continuum discretized in momentum space. They observe that there are no straightforward boundary conditions nor propensity rules on energy sharing and direction of emission, on the contrary, a very complex relation between the angular emission patterns of electron and positron emerged from Scheid's theoretical calculations for collision systems 10 AGeV $Au^{79+}$ + $Au^{79+}$. One of the quite interesting findings is that leptons are emitted with mean energies $<E_{lepton}> \approx < 1MeV$ and that even at these high collision energies only a small fraction of the emission cross sections is seen with high energy leptons >1MeV.

For further discussions of essential and indispensable spectrometer parameters we focus on two surprising and highly remarkable results reported by Scheid's group, which are playing a decisive role in our further considerations: the angular distribution of positrons from the above reaction (integrated over all electron energies and electron emission directions) exhibits two peaks in the forward and backward direction at around 50 and around 130 degree, respectively, superimposed over a near isotropic distribution (see fig. 4a) [ 28, 29].

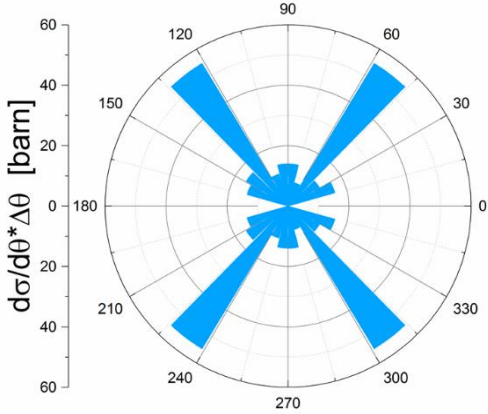

*Fig. 4a. Angular distribution of positrons in 10 AGeV $Au^{79+}$ + $Au^{79+}$ integrated over all energies and directions of the coincident electron [28, 29], discretization of momentum space results in angular bins of $\Delta\theta=15^0$ .*

An even more remarkable result was found for the distribution of difference angles $\theta_{e+,e-}$ for the emitted electron and positron of a pair. For free-free electron- positron pairs in the investigated system it could be established that there is definitely no evidence for a preferred back to back emission; their calculations rather exhibit a very broad distribution. Even lepton pairs with near vanishing difference angles are produced with a significant probability. A propensity for difference angles in a very broad distribution around 90 degree is observed in fig.4b; we also note a small but non-vanishing share of pairs to appear with back to back emission. Tenzer et al. also showed that only for transverse emission of the electron the corresponding positron is emitted near $160^0$ with respect to the electron polar emission angle, whereas for other polar emission angles of the electron no such propensity is found [28, 29].

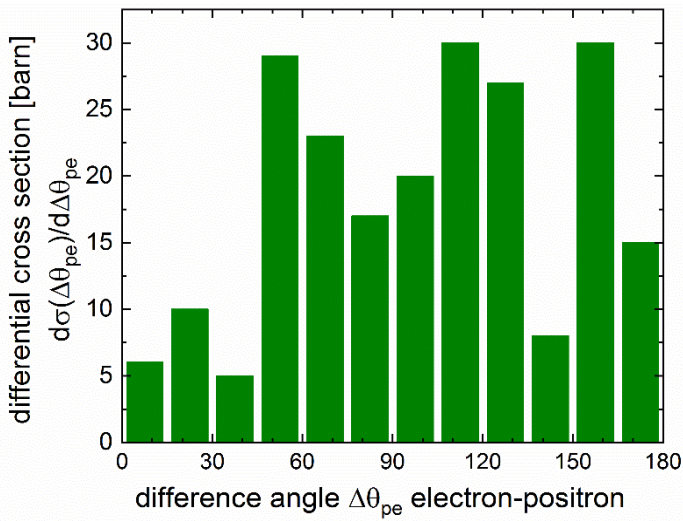

*Fig. 4b. Distribution of difference angles $\Delta\theta_{e^+,e^-}$ in 10 AGeV $Au^{79+}$ + $Au^{79+}$ for arbitrary energy of leptons exhibiting that in ion-atom collisions there is no clear preference for back to back emission of the coincident leptons [28, 29]. Note, that a given difference angle $\Delta\theta_{e^+,e^-}$ for a lepton pair does not predict an absolute orientation angle $\theta_{e^-}$ or $\theta_{e^+}$ for a lepton of the pair with respect to the beam axis.*

Similarly, it had also been shown by H. Bhabha that electrons and positrons created in swift collisions of ions and atoms [25] emerge at angles independent of each other with a propensity for small pseudorapidity, i.e. perpendicular to the direction of the projectile beam, and a rather steep decrease with the lepton energy. Another differential cross section for a near symmetric and very heavy collision system was studied by D. Ionescu in his Habilitation (18, 36 see fig 5 which only covers a small section of the entire available phase space, with a rather high kinetic energy of both leptons).

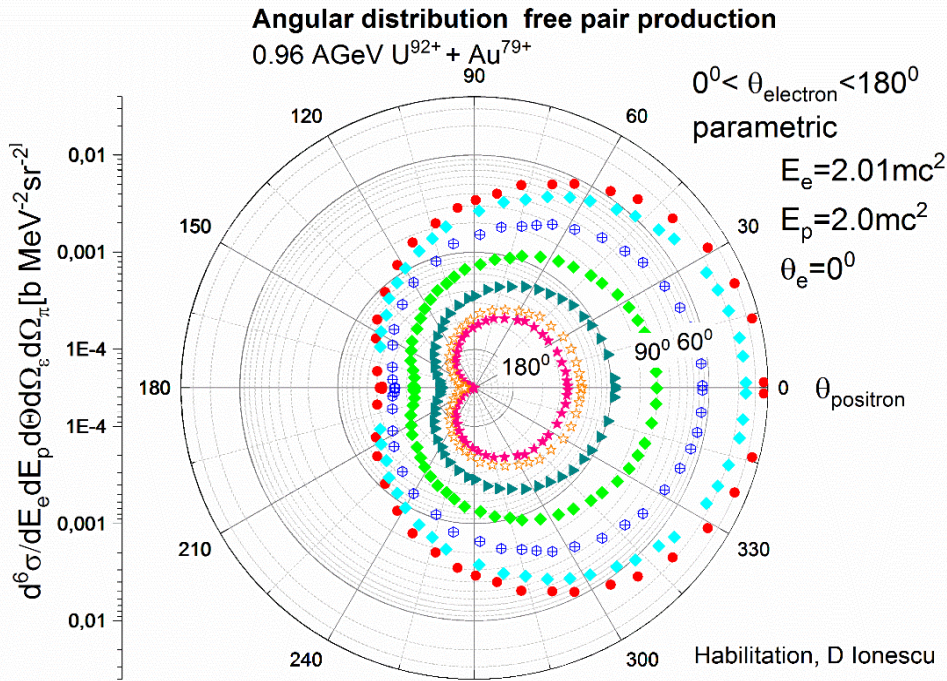

*Fig. 5. Angular distribution of free electron positron pairs as function of positron angle with electron angle $\theta_e$ as parameter, for lepton energies $E_{electron}=2.01mc^2$ and $E_{positron}=2mc^2$, respectively; Sommerfeld-Maue wavefunctions for the free electron positron states are used in this perturbative calculation [18].*

These first theoretical benchmark calculations show that a detailed experimental study of the correlated vector momenta of coincidently emitted electrons and positrons in free-free pair production is urgently needed in order to execute a comprehensive test of the current most advanced theories.

Due to the formidable experimental obstacles there are up to this time no experimental full fourfold differential cross sections (4DCS) for free-free pair production $\frac{d^4\sigma}{dE_{e^+}d\Omega_{e^+}dE_{e^-}d\Omega_{e^-}}$, or $\frac{d^4\sigma}{dp_{e^+}d\Omega_{e^+}dp_{e^-}d\Omega_{e^-}}$ respectively, in the literature, fully differential in energy/momentum and emission direction of both coincident leptons. Such 4DCSs evidently would present the most complete information on the collision dynamics and would simultaneously constitute the most stringent test of theories. Similarly, on the theoretical side it is the immense computing power required for a true unrestricted two-lepton continuum process which has purportedly kept theorists from actually calculating full 4DCS for free-free pair production but focus rather on some selected benchmarks and a rather coarse momentum and angular resolution, as shown in Scheid's work.

Under these circumstances it appears attractive to first address for an experimental study threefold differential cross sections

$$3\text{DCS}(e^+) = \frac{d^3\sigma}{dE_{e^+}d\Omega_{e^+}dE_{e^-}} \quad \text{and } 3\text{DCS}(e^-) = \frac{d^3\sigma}{dE_{e^-}d\Omega_{e^-}dE_{e^+}}$$

or the corresponding 3DCS in momentum space, where one has integrated over the emission direction of one member of the lepton pair; these 3DCSs may serve as a first, moderately restricted though urgently needed experimental test of advanced theories.

Our present study serves to show the feasibility of a spectrometer for an experiment detecting with a large solid angle for both leptons in coincidence electrons and positrons from free-free pairs produced in relativistic ion-atom collisions and derive selected 3DCSs and partially 4DCSs for heavy ion atom collisions systems. For experimental investigations using magnetic spectrometers the momentum space is the natural frame and thus in the next sections all differential cross sections will be provided and discussed in momentum space, e.g. the 4DCS $\frac{d^4\sigma}{dp_{e^+}d\Omega_{e^+}dp_{e^-}d\Omega_{e^-}}$ and corresponding 3DCSs $3\text{DCSp}(e^+) = \frac{d^3\sigma}{dp_{e^+}d\Omega_{e^+}dp_{e^-}}$ and $3\text{DCSp}(e^-) = \frac{d^3\sigma}{dEp_{e^-}d\Omega_{e^-}dp_{e^+}}$., for details see section 6d and for the corresponding transformations appendix 2.

3. **The high energy storage ring HESR**

The relativistic high energy storage ring for ions and antiprotons HESR was originally conceived to store and cool antiprotons for the PANDA collaboration at FAIR [58]; but very soon additional extended ion-optical calculations [57, 59, 60] proved the HESR to be a highly promising facility for studies of atomic collision dynamics and few-electron high-Z atomic spectroscopy in the realm of near relativistic heavy-ion atom collisions. The collision energy range offered by the HESR is unique in the world and is particularly interesting for Coulomb dominated free-free pair production (see fig 6a).

In the following we will only list a few essential parameters of this unique storage ring, which are useful for our further discussions, but we refer the inclined reader to the literature for more detailed information [57-60].

The HESR with a magnetic rigidity ranging from 5 to 50 Tm has a circumference of 550 m and can store highly charged ions up to $U^{92+}$ with energies between 170 AMeV and 5AGeV, corresponding to $\gamma_{max} \approx 6$.

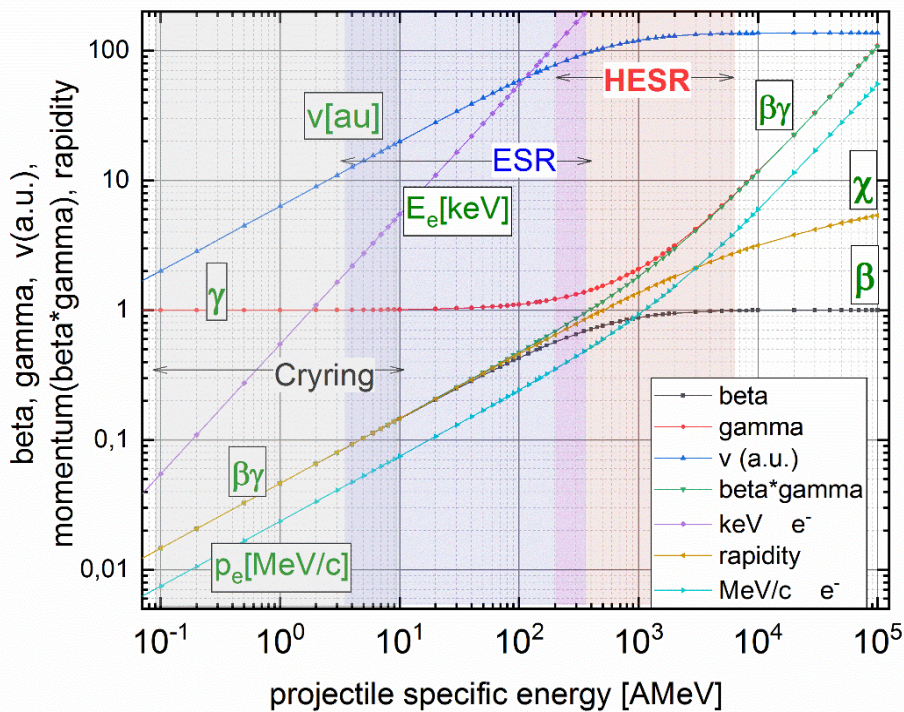

*Fig 6a. Comparison of collision energy range of the future HESR storage ring with current Storage rings ESR and Cryring at GSI. It is nicely visible that the HESR extends from near relativistic regime into the pure relativistic regime where $\beta\gamma$ is clearly dominated by $\gamma$.*

Beams from synchrotrons, the current SIS18, or the future SIS100, will be injected at 740AMeV energy and can be decelerated and/or accelerated to the energy desired by the respective experiments. Stochastic cooling and electron cooling will be possible between 740AMeV and 5AGeV in order to provide a beam with a momentum spread $\Delta p/p$ below $10^{-4}$. The storage ring is equipped with two long straight sections, one for the PANDA detector, and the other for a variety of experimental installations including the electron cooler. At two alternative locations indicated in fig. 6 a supersonic jet may be implemented [60] for experimental installations ranging from high resolution x ray crystal spectrometers and calorimeters to magnetic pair spectrometers.

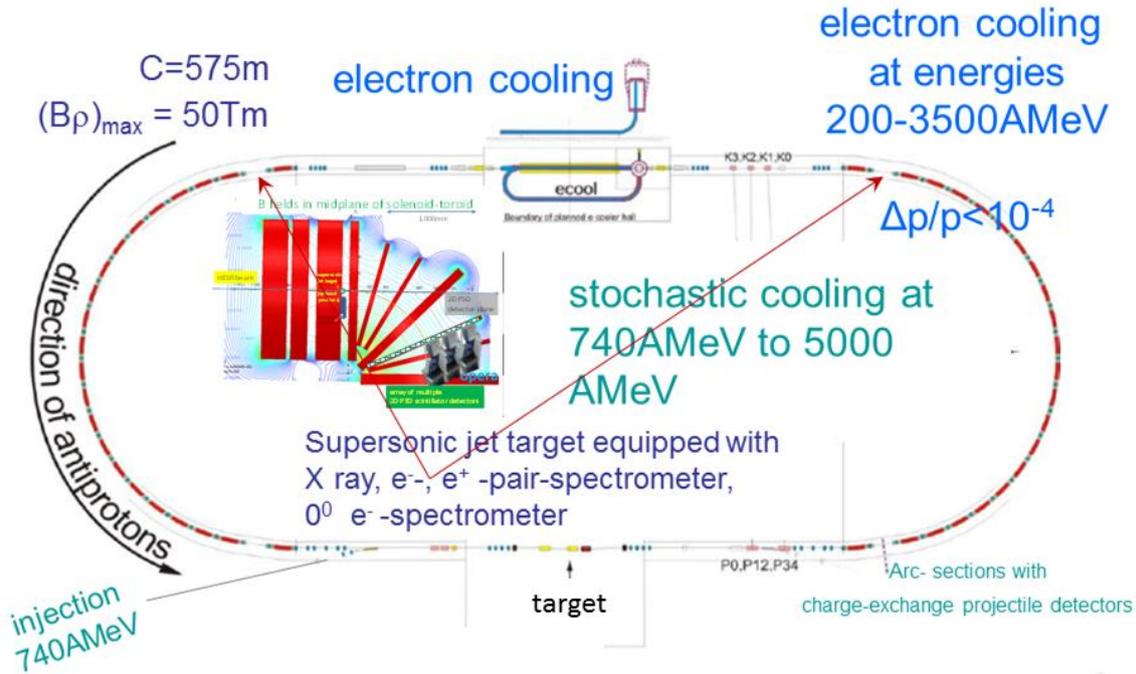

*Fig. 6b. The relativistic high energy storage ring HESR with selected installations, the two options for the installation of the magnetic pair spectrometer in the arc section immediately in front of and following the electron cooler, respectively are indicated by red arrows, for details see [57, 59, 60] .*

From extensive electron-optical and ion-optical calculations which were performed when implementing our magnetic $0^0$ electron spectrometer [124] in the supersonic jet target zone of the current ESR storage ring we know that the effect of the magnetic field of the spectrometer on the coasting ion beam could completely be compensated by two small magnetic steerers in the ring lattice. Corresponding calculations for the toroidal spectrometer in the HESR are in progress.

### 4. Kinematics of free-free pair production in an emitter of near relativistic collisions velocity

The kinematic maximum in the positron spectra reported by Belkacem et al. [34, 40] and the calculations of Scheid et al.[28, 29] demonstrate that the continuum of the two emitted leptons for free-free pair production is in the emitter frame dominated  by leptons of low kinetic energies. It thus appears indicated to revisit for relativistic collision systems the spectral features which are expected for the 2-lepton continua in the laboratory and which can be generated by the highly diverse theoretically predicted emission patterns of a free-free pair in the projectile emitter frame. For near-relativistic collision systems with $(\gamma-1)>>0$ the Lorentz transformation from the emitter- to the laboratory frame complicates the easy graphic derivation of kinematic relations so well established from Newton diagrams in non-relativistic kinematics. A relativistically correct graphic representation[71 - 74] does nevertheless also exhibit all the important material features, e.g. a maximum laboratory observation angle $\theta_{max}$, which is suitably calculated following $sin\theta_{max} = \frac{\beta'\gamma'}{\beta\gamma}$ ($\beta'$, $\gamma'$ for the lepton in the emitter frame, $\beta$, $\gamma$ for the emitter in the laboratory frame) for a slow lepton emitted by a fast projectile and also the momenta of leptons in the laboratory as function of laboratory observation angle; this instructive and nearly indispensable tool [74] had a high practical value before the advent of computers, today it retains a very desirable didactic value (see fig. 7), though.

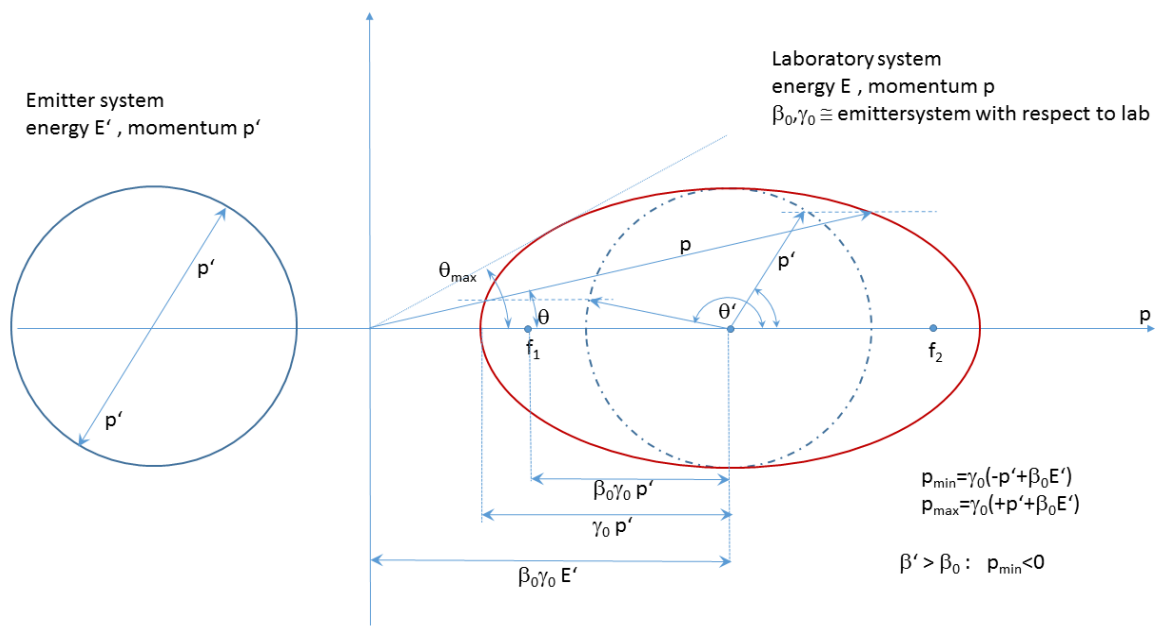

*Fig. 7. Illustration of useful kinematic relations for momenta in emitter and laboratory frame in relativistic kinematics*

In order to illustrate the features adequate for a useful pair spectrometer we consider as a first step the kinetic energy and the angular distribution in the laboratory expected for a hypothetical scenario at the future HESR: we begin with leptons emitted with fixed energy $E_{kin}=500$ keV in the projectile frame for 5.1AGeV projectile energy.

Using the relations

$$E_{e,kin}^{lab} = \gamma \left( E_{e,kin}^{proj} + \beta \sqrt{\left(E_{e,kin}^{proj}\right)^2 + 2E_{e,kin}^{proj} * mc^2} \cos \theta_{proj} \right) +$$

$$(\gamma - 1)mc^2 \qquad \qquad (4a)$$

or in a more compact form using momenta

$$E_{e,kin}^{lab} = (\gamma \gamma_e^{proj}(1 + \beta \beta_e^{proj} \cos \theta_{proj}) - 1) \ mc^2 \qquad (4b)$$

for the dependence of the kinetic energy of the lepton in the laboratory frame $E_{e,kin}^{lab}$

on energy and emission angle in the emitter frame, $E_{e,kin}^{proj}$ and $\theta_{proj}$, respectively,

and

$$\sin \theta = \frac{\sin \theta_{proj}}{\sqrt{\gamma^2 (cos \theta_{proj} + \frac{\beta}{\beta_e^{proj}})^2 + sin^2 \theta_{proj}}} \qquad \text{or} \qquad (5a)$$

$$tan \theta = \frac{\beta_e^{proj} sin \theta_{proj}}{\gamma(\beta_e^{proj} cos \theta_{proj} + \beta_0)} \qquad , \qquad (5b)$$

respectively, for the relation between emitter frame emission angle $\theta_{proj}$ and laboratory observation angle $\theta$ we arrive at the instructive graphical relation in fig. 8.

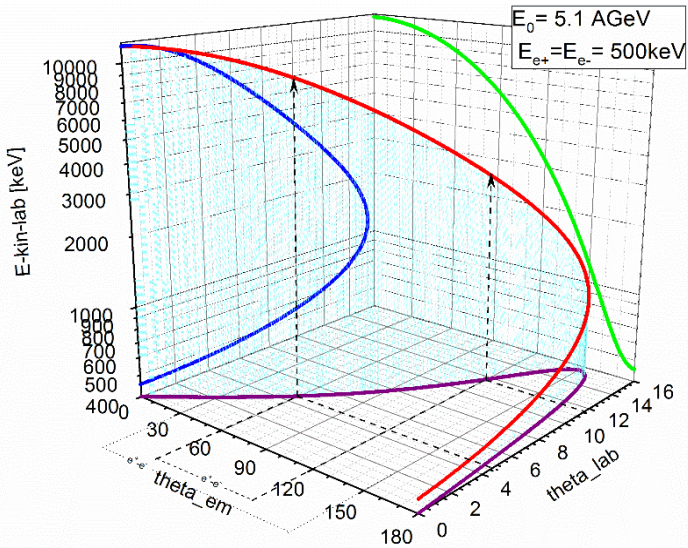

*Fig. 8. Kinematical relations for leptons emitted by a fast emitter, e.g during pair creation: a hypothetical e⁺ - e⁻ pair with $E_{e,kin}^{proj}$ = 500 keV for each leptons is emitted in a moving frame of 5.1 AGeV (γ≈6). If such a pair of leptons is assumed to be emitted back to back e.g. at $60^0$ and $120^0$ in the emitter frame (dashed lines) it will appear in the laboratory as coincident pair with 9 MeV at $4.5^0$ and 3.3 MeV at $11.5^0$, respectively.*

We observe that the lepton emission in the fast moving emitter produces as a general feature the well-known narrow forward cone of leptons in the laboratory with a wide range of kinetic energies. It is apparent that even very strict kinematical conditions assumed for the emission characteristics of a hypothetical lepton pair in the emitter frame will still encompass a wide range of kinetic energies of leptons, e.g. between 400keV and 20MeV which will have to be covered by actual lepton detectors, as illustrated in the example in fig. 8. However, only when assuming back-to-back emission of both leptons (and with a further restriction of both leptons emitted with the same kinetic energy in the emitter frame), it is straightforward to determine from coincident measurement of the laboratory energies of the leptons the emitter frame energy and emission angle, as is illustrated in fig. 9. On the other hand, a hypothetical assumption of equal energies in the emitter frame with no assumption on emission angles, though, will not result in a unique complete identification of primordial emission characteristics.

For ion-atom collisions, however, the open phase space spanned by the two leptons of a free-free pair in the continuum is very large and is practically not constrained by any kinematic conservation rules, such as back to back emission of leptons in the emitter frame. Theory [24-25, 28-29, 36] has predicted that the two leptons of a coincident pair may be indeed emitted with a nearly unrestricted, very wide range of difference angles (in the emitter frame) and with no strict rules applying, neither for respective emission directions, nor for energy sharing. We illustrate for typical lepton energies in the emitter frame for 5.1AGeV Au + Au, i.e. 200 keV, 500 keV, 1MeV and 2 MeV, the kinematic relations resulting in the laboratory for emission of leptons over a wide range of kinetic energies, in fig. 9; this illustrates the very large phase space which needs to be covered by a spectrometer when both leptons of the pair are to be detected in coincidence.

We wish to draw attention to two interesting features:

i) for lepton emission up to around $90^0$ in the projectile frame the relation between emitter frame emission angle $\theta_{proj}$ and laboratory observation angle $\theta$ exhibits only a very weak dependence on the primordial kinetic energy of the lepton. For a hypothetically preferred transverse emission at $\pm90^0$ of the free-free pair in the emitter frame, and equal energy sharing for the leptons and kinetic energies $E_{e,kin}^{proj}$ of .5, 1 and 2 MeV, respectively, both leptons would appear under laboratory observation angles quite close to $\theta=8^0$ for all corresponding laboratory kinetic energies, 6, 10 and 17 MeV.

ii) near the maximum laboratory scattering angle, i.e. at backward emission in the emitter frame, the laboratory energy of the leptons is rather low as a consequence of kinematic focussing and the relation $E_{lab} \leftrightarrow E_{emitterframe}$ for the leptons is inverted. Leptons from $(e^-, e^+)$-free-free pair production with a low energy in the laboratory are all originating from high energy leptons emitted in the backward direction in the emitter frame due to their production in a fast moving emitter. This has a considerable implication for the observation of leptons and the spectrometer design: the transverse momentum of the leptons which determines the magnitude of the gyroradius in a magnetic field is rather small. This will be discussed in more detail in the context of the lepton trajectories.

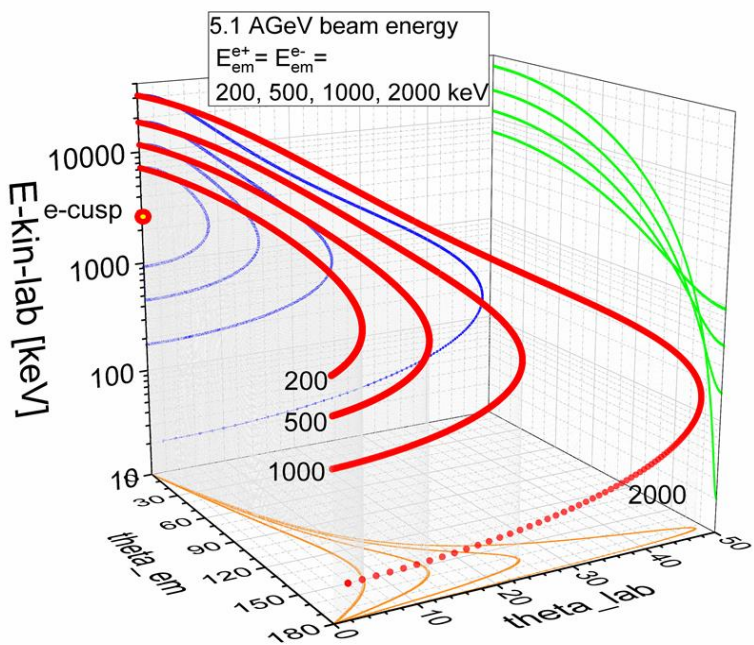

*Fig. 9. Kinematical relations for leptons emitted by a fast projectile, for lepton energies 200keV, 500keV, 1MeV and 2MeV in the emitter frame. Note the wide range of laboratory kinetic energies of the leptons extending beyond 30MeV in the forward direction. The diagram shows that the large laboratory observation angles correspond to small laboratory energies for backward emission in the emitter frame $\theta_{em} \geq 150^0$.*

The electron branch of the toroid spectrometer will see high-energy electrons mainly from three processes:

a) δ-electrons from target ionization processes (i.e. bound state electrons of the target ionized by the time dependent perturbation by the projectile potential). The momentum transfer $\langle q \rangle = \frac{E^b + E^{cont}}{hv}$ ($E^b$ initial state binding energy, $E^{cont}$ final state continuum energy of electron, v collision velocity) necessary to generate high energy (≈MeV) δ-electrons implies that these electrons originate mostly from very high momentum components of the Compton profile of the initially bound target electron [75] if they appear at MeV laboratory energies. Corresponding cross sections are very small.

b) As well ionization of the projectile via electron loss to continuum (ELC) may contribute, when a non-bare projectile carries electrons into the collision; these electrons are easily identified: they appear in coincidence with a projectile whose final charge state is increased by n for n-fold projectile ionization and exhibit a narrow cusp at $\theta \approx 0^0$ with $v_e \approx v_{proj}$ [124].

c) Electrons from free-free pair creation mainly under consideration here will see some overlap in the same spectral region as the two above ionization channels a) and b). For free-free pair creation according to eq (1a) the projectile does not experience a charge exchange [36].

For the positron branch in the spectrometer positrons from weak interaction nuclear processes may arise besides those of atomic origin; however, the lifetime of the corresponding emitter states will lead to negligible decay before the ion passes the spectrometer entrance.

We illustrate with this example, that due to the lack of strong kinematic boundary conditions in lepton pairs produced in ion atom collisions true electron-positron coincidences may be located anywhere in the very extended 2-lepton phase space given by the very large range of $E_{e,kin}^{lab}$ and $\theta_{lab}$ (see figs. 9 and 10a,b). In order to achieve reasonable electron-positron coincidence count rates in an experiment any new spectrometers will have to cover simultaneously a large range of kinetic energy $\Delta E_{e,kin}^{lab}$, or electron momenta, respectively, and solid angle $d\Omega_{lab}$ for the observation angles $\theta_{lab}$ without sacrificing the desired energy and angular resolution. For a lepton pair originating in a 5.1AGeV collision each lepton's phase space is depicted in fig. 10a.

The largest angles $\theta_{lab}$ are associated with low lepton momenta and will thus be mapped onto small perpendicular momenta/radii. It is the electron and positron from near forward emission at small $\theta_{lab}$ which will dominate the events at perpendicular momenta.

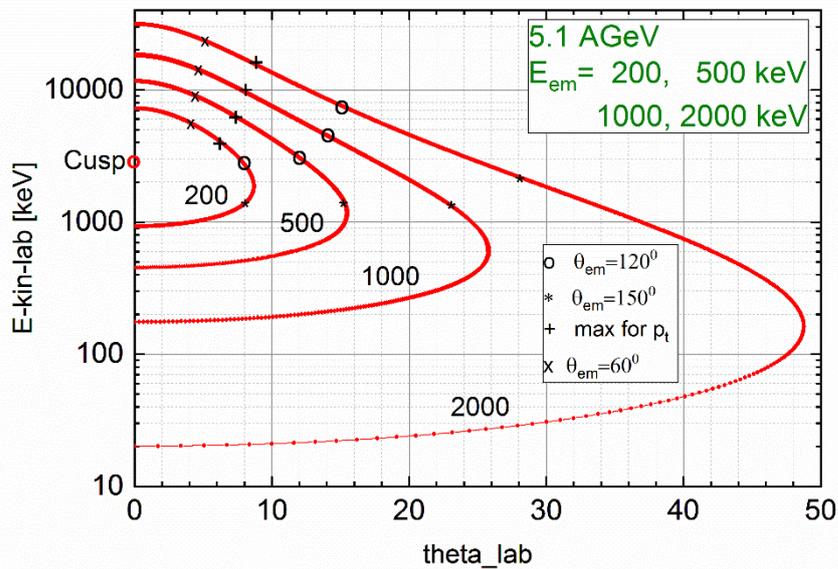

*Fig. 10a. Laboratory phase space for leptons emitted by a fast projectile: kinetic energy −observation angle range for free-free pair production for the case of projectiles with a specific kinetic energy 5.1 AGeV and a sample of four emitter frame energies of a lepton. For a lepton (e.g. e+) with a given energy the laboratory observation angles of its coincidently emitted lepton (corresponding e-) can cover a wide range according to Scheid [ 28-29] and the electron principally may appear anywhere in this graph with no a priori reasonable limit for the range of kinetic energy and emission direction provided by theory. It is apparent that a combination of two independent position dispersive instruments, even when equipped with 2D-PSD lepton detectors will have prohibitively small combined coincidence efficiency. The ticks for emitter frame angles 60⁰, 120⁰ and 150⁰ illustrate a favourable strong kinematic forward emission in the laboratory frame. Virtually the entire forward hemisphere is mapped into $\theta_{lab} \leq 10^0$; even at $E_{emitter}$= 2MeV the emitter frame emission up to 120⁰ is mapped into lab angles < 15⁰. For every emitter frame energy of a lepton the cross indicates on the abscissa the laboratory angle where the transverse momentum has its maximum. It is apparent that for the*

*leptons the large laboratory emission angles correspond to a small laboratory momenta and corresponding laboratory kinetic energy.*

Even though the measured laboratory kinetic energy combined with the laboratory observation angle of a detected lepton permits to uniquely identify its primordial vector momentum in the emitter frame it is apparent that traditional spectrometers with a meaningful energy and angular resolution will have a prohibitively small coincidence efficiency, due to their type of position dispersion and small solid angle.

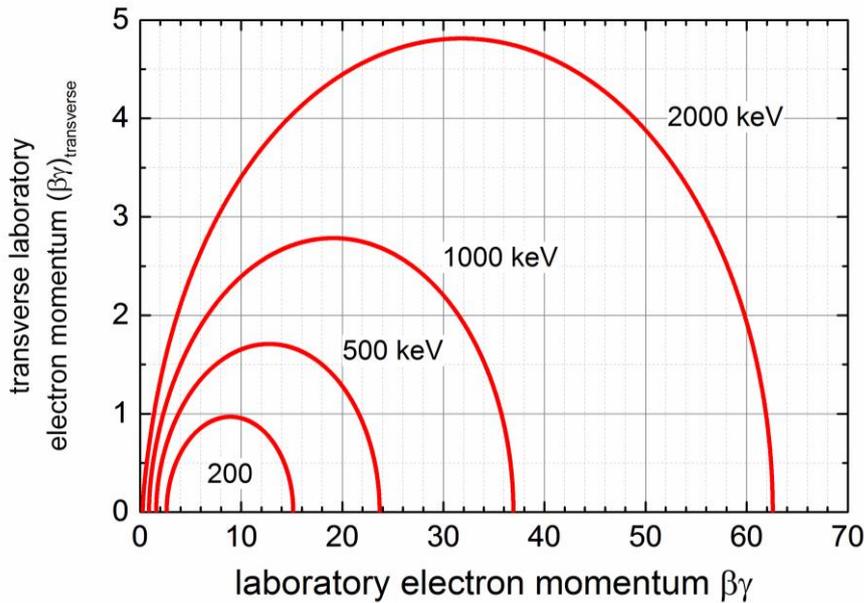

*Fig. 10b. Relation of $p_{transverse}$ and $p_{tot} = \beta\gamma$ in the laboratory of leptons emitted by 5.1AGeV projectiles. A lepton is uniquely and completely characterized by $p_{transverse}$ and $p_{tot}$ and thus its vector momentum in the emitter frame can be reconstructed. For emitter frame kinetic energies of 2MeV and 1MeV, respectively, leptons with low laboratory momenta $p = \beta\gamma < 30$ and $\beta\gamma < 20$, respectively, present small fractions of the total cross section with emitter frame backward emission cones of $10^0$ and $15^0$, respectively around the backward direction (see also fig 9).Note that the apex of each curve, the largest $p_{transverse}$, does not correspond to the largest assumed $\theta_{lab}$.*

The laboratory emitter angle $\theta_{lab}$ determines for a lepton's momentum $\beta\gamma$ its transverse component $(\beta\gamma)_{transverse}$, which will in the small angle paraxial approximation permit to derive an approximate radius of gyration $\varrho$ in a magnetic field B via $B\varrho[Tm] = 1.7045 * 10^{-3}(\beta\gamma)_{transverse}$ This results for perpendicular momenta $(\beta\gamma)_{transverse}$=4.8 ($\leftrightarrow$ $E_{kin}$= 2 MeV) in a radius of gyration of $\varrho \approx 8$ cm for a magnetic field B=1000G [and gives equiv. $(\beta\gamma) = 0.48, i.e. 55\ keV$ in a $100G$ field ]. All leptons whose momentum $(\beta\gamma)_{transverse}$ is smaller than $586.7 B\varrho[Tm]$ are detected within a circular area with diameter 2 $\varrho$ in the detector plane. (Note for electron momenta $p_e$[MeV/c]=0.511$\beta\gamma$ and in atomic units $p_e$[a.u.]= $\beta\gamma/\alpha$)

Even position dispersive spectrometers (e.g. magnetic sector spectrometers [75-78,137]) with 2D position sensitive lepton detectors can only partially improve the overall coincidence detection efficiency as the energy or momentum window both instruments see at any given time is small compared to the very large range of lepton momentum/energy and emission angle which needs to be covered for acquiring differential cross sections over a meaningful dynamic range.

### 5. The Magnetic Toroidal Sector

A major stimulus and valuable guide in pursuing a study of electron-optical properties of a magnetic toroidal sector for lepton pairs originating from relativistic heavy ion atom collisions has been the TORI spectrometer conceived for experiments on adiabatic positron production in superheavy transient quasimolecules [78-80], e.g. collision energies $\geq$ 5.9AMeV for $U^{q+}$+Pb, close to the Coulomb barrier; TORI exploits a fundamental property of a toroidal magnetic field, separating particles of opposite charge perpendicular to the bend plane of the toroid; TORI was developed and operated in a series of experiments by the Kankeleit group at the UNILAC. In the TORI spectrometer electrons and positrons emitted by transient superheavy quasimolecules [64, 76, 77] created in a solid high Z target during adiabatic collisions with incident high Z heavy ions are guided in the first section by a toroidal +90$^0$ magnetic sector from the target to an intermediate plane where a baffle intercepts the electrons, which follow at this location trajectories geometrically well separated from those of the positrons. The positrons pass this midplane and enter a second section, point mirror symmetric to the first sector with respect to the midplane; then, in this second 90$^0$ toroidal sector the positrons are guided back onto the optical axis at the end of this section and there onto an array of PSD detectors [84-87,91].

The excellent suppression of background-electrons arising from the solid target combined with a very high overall efficiency for positron detection are reason for the success of these experiments studying positron production (and delta-electron production as well, in a corresponding configuration [75]) in transient superheavy quasimolecules created in adiabatic collisions of very heavy ions and heavy atoms [76, 77].

The fundamental property of a toroidal magnetic field, to disperse particles of opposite charge perpendicular to B and to ∇B into opposite directions and thus allowing to guide them to spatially isolated PSDs, appears to be also particularly suitable for experiments investigating free-free-pair production planned at the future HESR storage ring at GSI/FAIR; here, for free-free pair production, in a significant step beyond positron-only spectroscopy as used in TORI, vector momenta of the positron and the corresponding electron must be determined simultaneously. For accomplishing this class of experiments we require, however, just one toroidal sector, because such a toroidal B-field configuration intrinsically permits to simultaneously analyse the vector momenta of both leptons of the pair in coincidence at a location of optimal separation (see fig 11), when using position sensitive detectors for both, electrons and positrons in the detector plane.

## B fields in midplane of solenoid-toroid

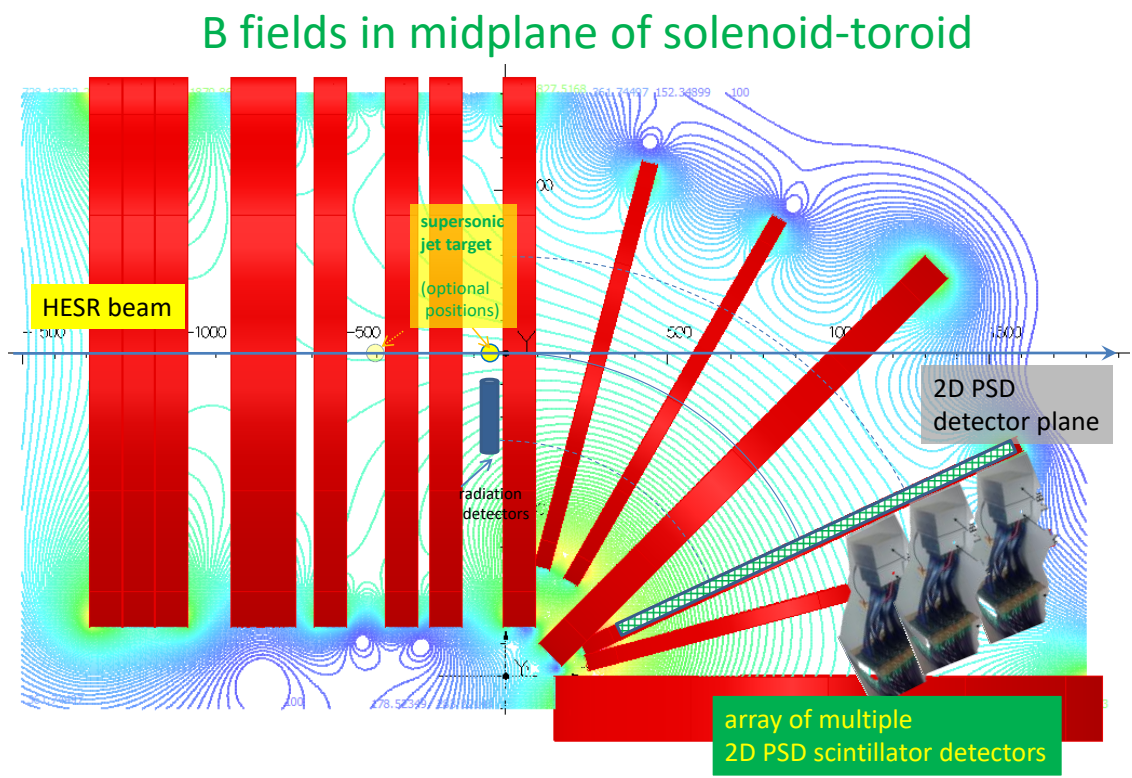

*Fig. 11. Cut through the mid-plane of the coil assembly of a toroidal magnetic sector spectrometer for simultaneous electron and positron spectroscopy at the supersonic gas jet target in the HESR storage ring. The centre line of the toroid is depicted as a solid blue circular arc- sector extending to coil 5 (location of 2D-PSD detectors). Two dashed lines indicate locations of ±300mm from the centreline in the mid plane. The coil spacing in the solenoidal part reflects the option to facilitate access to the target zone for other detectors, e.g. x-ray and other photon detectors and as well 2D-PSD recoil ion detectors. We assume for illustration of the geometry a local cylindrical coordinate system (ρ,φ, z) with its origin at the centre of curvature of the toroid, here*

*we have ρ=1000mm and φ=0 at the entrance of the toroid. The corresponding z-axis is perpendicular to the centre plane, +z pointing out of plane. In this coordinate system the magnetic induction B in the toroidal section - when generated by an effective meridional current in the coils - is ideally tangential $\vec{B} = \frac{A(i)}{R} \vec{e}_\varphi$, where R=distance from the origin to the location of the field measurement.*

For implementation of a magnetic toroidal spectrometer inside a storage ring the boundary conditions of operation - a minimum perturbation to the coasting particle beam passing through several coils of the set (and subsequent compensation of this perturbation to the coasting beam using additional beam optical elements in the ring lattice downstream of the toroid) - impose significant limitations on the geometry for any configuration of the magnetic coils which shall produce the desired toroidal magnetic field; obviously a single contiguous toroidal coil, although desirable, is not compatible with boundary conditions characteristic for operation in a storage ring. (In storage rings the injection orbit of ions is in general quite distinct [59, 60] from the orbit chosen for the data taking cycle, which may only commence after cooling and optional cycles for deceleration to the desired collision energy – thus imposing strict conditions on all in-ring spectrometer designs.) The transverse opening required by the storage ring beam optics of min 200mm H and 100mm V combined with the desire of minimizing the distortion of the homogeneity of the toroidal magnetic field configuration due to separated coils producing the toroidal B field entails a toroidal radius R of approximately 1000mm and corresponding coil sizes.

In order to maintain the flexibility for complementing the lepton pair spectrometer with a reaction microscope [88] containing a full 2D position sensitive detector in the target zone in the straight section [94] our design of up to 10 coils on an arc radius of 1000mm for generation of the toroidal field is complemented by up to 7 coils for a solenoidal section upstream of the supersonic jet target zone.

In an ideal magnetic toroid the magnetic induction B is tangential $\vec{B} = \frac{A(i)}{R} \vec{e}_\varphi$ assuming that the current in the coil is purely meridional, i.e. has no component along $\vec{e}_\varphi$. This normally implies a contiguous toroidal coil for generating this field. The boundary conditions for in-ring spectrometers in the HESR storage ring, however, prohibit this solution. We have thus studied the magnetic fields produced by various coil configurations. In fig. 12 and 13 we show - using calculations of a corresponding coil configuration shown in fig. 11 developed with TOSCA/OPERA-3D [81] - that the current configuration of 10+7 coils permits to generate a toroidal B-field with approximately

ΔB/B≈±1*10⁻³ deviation from the ideal field over the centre arc and a region of a cross section of more than ±200 x 200 mm around the centre arc perpendicular to the optical centre line from the target zone all the way to a detector plane. We investigate detector planes positioned between φ=45⁰ and φ=83⁰, depending on the relative spatial separations of the leptons of different energies desired; in fig. 11 a detector array in position at φ=66⁰ (i.e. coil 5 of the toroid section) is schematically depicted.

An array of multiple 2D PSD-scintillators [85-87] with energy resolution down to ΔE/E<0.01 provides a position resolution down to ≤ 0.5 mm; this good energy resolution for incident leptons adds for incident leptons highly useful redundancies in the correct lepton identification.

For experiment configurations requiring detection of leptons over a wide range of low kinetic energies in the laboratory frame, including energies close to threshold, and simultaneously of very high energy leptons close to the beam direction, e.g. Cusp or binary encounter electrons, an option for a detector plane at Θ=45⁰ will cover all energies for leptons from threshold to very high energies at forward emission, however with a slightly decreased momentum dispersion for leptons.

Additional simulation runs verified that optional jet target positions (desirable for operation in conjunction with a reaction microscope) extending up to -500mm upstream (with respect to the position (1000,0,0)) in the solenoidal section do not visibly deteriorate lepton momentum dispersion in the detector planes.

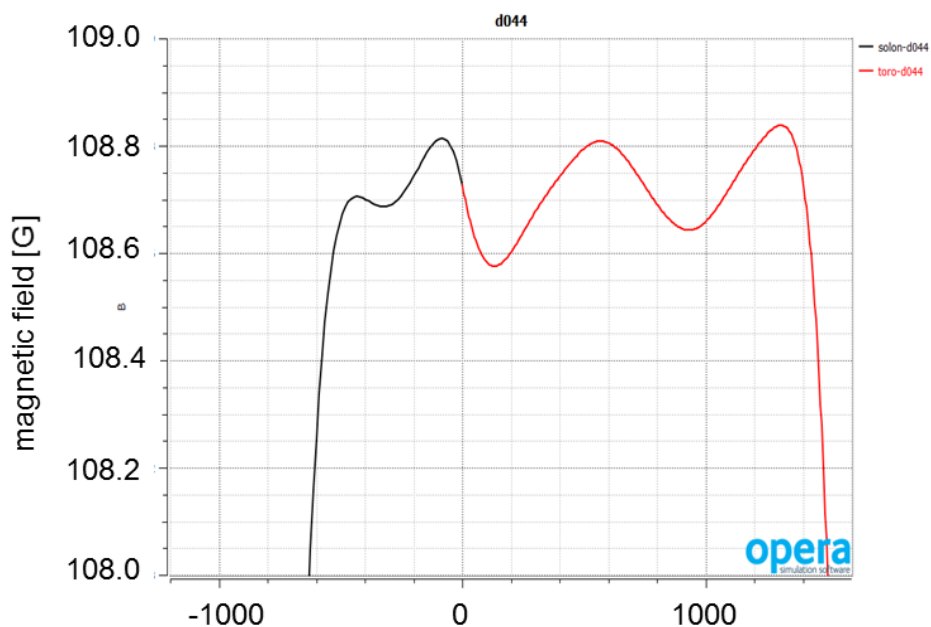

*Fig. 12. Magnetic B-field uniformity: B-Field along centre line of solenoid (coil 1 to 3) and centre arc of toroid calculated using OPERA-3D for the coils shown in fig. 10. The abscissa 0 corresponds to the entrance of the first coil of the toroidal sector (=coil 4 counted from left), the detector position depicted in fig 10 (=coil 5 of toroidal section) is located at 1187mm on the arc. The deviation from the mean value does not exceed $\Delta B/B \approx \pm 1*10^{-3}$ in the region of interest.*

All magnetic field calculations were executed using the TOSCA section of the code OPERA_3D_10.5 which besides calculating the magnetic field from solving partial differential equations with boundary conditions

$$\nabla \cdot \mu \nabla A - \nabla \cdot \mu \left( \int_{\Omega_J} \frac{J \times R}{|R|^3} d\Omega_J \right) = 0, \qquad (6)$$

with $\mu$ the permeability tensor and **A** the magnetic potential, also calculates and displays trajectories of charged particles in the calculated magnetic fields[81].

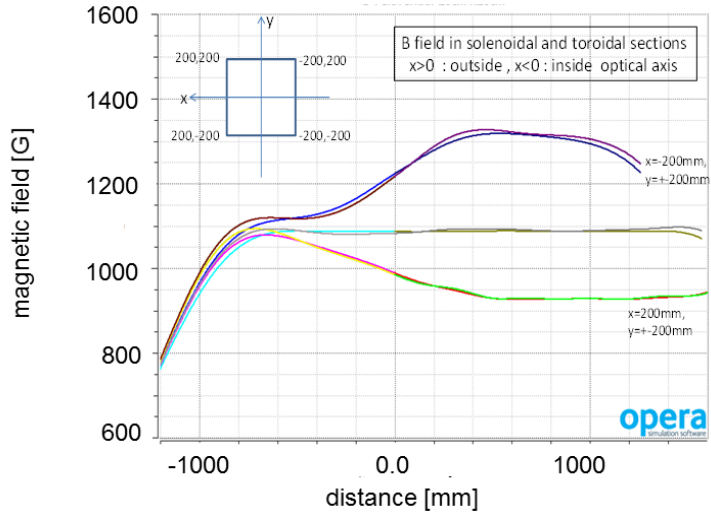

*Fig. 13. The B-field of the coil configuration exhibits a toroidal behaviour over a large volume: The magnetic field on various arc sections for an inner radius R=800mm, a centre radius R=1000mm and an outer radius R=1200mm are shown. This shows that over an area $\pm$200mm x $\pm$200mm throughout the toroid the B field exhibits a satisfactory toroidal behaviour; the relative deviations are particularly small not only on the centre line( as seen in fig. 12) but also in the important low field region outside of the centre line, where $R_{eff}$ is larger than the centre line radius R.*

We have previously studied in conjunction with kinematic complete experiments in ion-atom collisions in the few AMeV collision energy range the electron-optical properties of a toroidal branch combined with a reaction microscope [82], but with emphasis exclusively on very low energy electron spectroscopy.

### a) Toroidal coordinates

Before analyzing the lepton trajectories in solenoidal and toroidal fields it may be interesting to consider the choice of a suitable coordinate system. Given the toroidal geometry of the planned spectrometer we examine some useful properties of toroidal coordinates. For a toroidal geometry of the coil configuration with a purely meridional current the cylindrical coordinate system frequently used in calculations for a magnetically confined plasma is not optimal [95-97]; it is compelling to treat the magnetic fields and the lepton dynamics in toroidal coordinates [98,122,123] (we use the nomenclature chosen by Moon and Spencer [98]):

$$x = \frac{aSinh\eta\, Cos\psi}{Cosh\eta - Cos\theta} \qquad y = \frac{aSinh\eta\, Sin\psi}{Cosh\eta - Cos\theta} \qquad z = \frac{aSin\theta}{Cosh\eta - Cos\theta} \quad , \ 0 \leq \eta \leq \infty, \ -\pi \leq \theta \leq \pi, \ 0 \leq \psi \leq 2\pi,$$ and

*a* the scaling parameter for the foci (see figures 14a, b); the toroidal coordinates are a natural generalization of the bipolar coordinates $x_p = \frac{aSinh\eta}{Cosh\eta - Cos\theta}$ and $y_p = \frac{aSin\theta}{Cosh\eta - Cos\theta}$. For illustrative purposes we add two figures representing the corresponding coordinate surfaces.

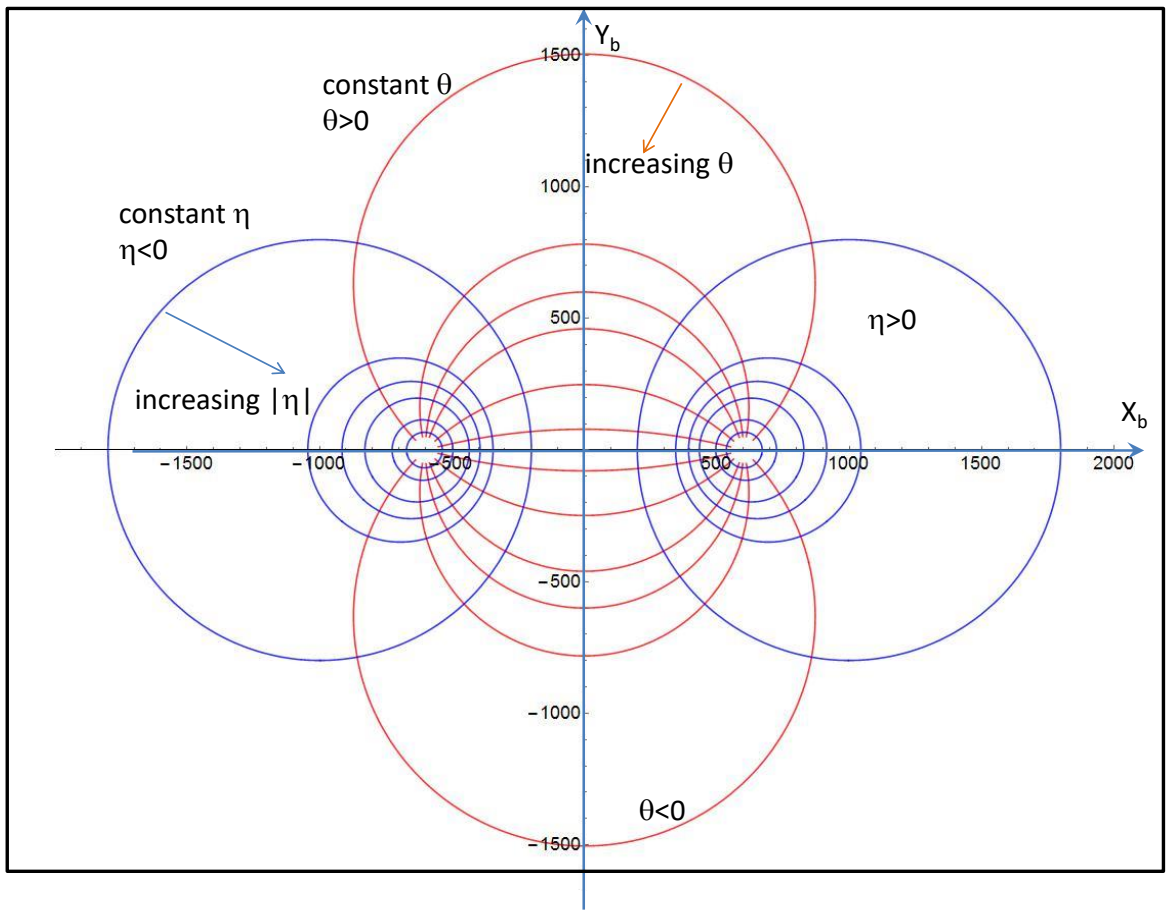

*Fig. 14a. Bipolar coordinates η and θ, -∞≤η≤∞, -π≤θ≤π. The curves of constant η (blue in fig. 14a) are non-intersecting, non-concentric circles around the foci -a and +a. For increasing |η| the radius of the circles decreases. The curves of constant θ are similarly non-concentric circles intersecting the foci –a and +a; for increasing |θ| the radius of*

*the circle decreases. The curves of constant $\eta$ and $\theta$, respectively, intersect at right angles; they turn, upon generalization, into toroidal coordinates with toroids and spheroids as coordinate surfaces.*

The fact that surfaces of constant $\eta$ are toroids makes toroidal coordinates very appropriate for investigation of magnetic fields generated by coils configured in toroidal geometry [99-102].

In fig 14a the parameters are selected such that the largest blue circles correspond to the actual coil geometry currently investigated with radius of curvature $\varrho$ =1000mm and coil-radius r=800mm and the foci calculated according to $a^2 = (\varrho - r)(\varrho + r)$ [100-102] to be at a=±600mm. The set of blue circles are the geometric location of points with a fixed ratio of distance to the foci and, respectively, the set of red circles are the geometric location of points under which the foci appear under a fixed angle; this system of orthogonally intersecting circles was elaborated on in antiquity as the circles of Apollonius [103].

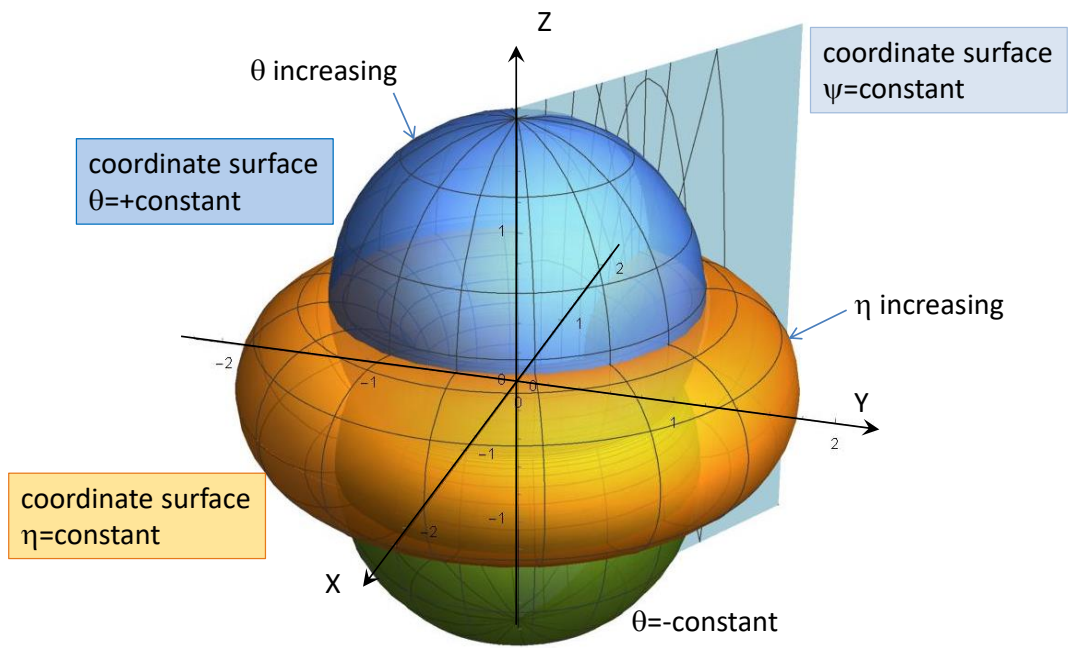

*Fig. 14b. Surfaces of constant $\eta$, $\theta$ and $\psi$ in toroidal coordinates following the nomenclature of Moon-Spencer. Note, that for increasing $\eta$ and $\theta$ the radii of the corresponding toroid and spheroid decreases (note corresponding error for toroidal coordinate directions in fig. 4.04 in Moon-Spencer [98]). For the actual magnetic coil configuration discussed here we have $\eta_0$ =0.7. In order to find the vector potential*

*inside the toroid, e.g. for the future calculation of the lepton trajectories in classical Hamiltonian equations of motion, one has accordingly to evaluate the vector potential for η≥0.7 (see appendix 1). Note, that in toroidal coordinates the magnetic vector potential has only one non-vanishing component $A_\eta \neq 0$, whereas $A_\psi = A_\vartheta = 0$ [101].*

### b) Helical Trajectories in a Solenoidal B-field

In order to better comprehend the character of lepton trajectories in a toroidal field we shall recall the main properties of the corresponding trajectories in a solenoidal magnetic B-field.

In a homogeneous solenoidal magnetic B-field a lepton with kinetic energy $E_{kin} = (\gamma - 1)m_0$ executes a helical trajectory around the B-field lines with a relativistic cyclotron frequency $f = \frac{qB}{2\pi\gamma m_0}$ as is illustrated in fig. 15. This corresponds to a revolution time $T = f^{-1} = \frac{2\pi\gamma m_0}{qB}$, a gyroradius $r_g = \frac{\beta_\perp m_0 c}{qB}$, where the perpendicular component of β is used in calculating $r_g$ and the wavelength of motion $\lambda_L = \frac{\beta_\parallel c}{f} = \frac{2\pi\gamma m_0 \beta_\parallel c}{qB}$. In the non-relativistic approximation γ=1.0 cyclotron frequency $f_{nr} = \frac{qB}{2\pi m_0}$ and revolution time $T_{nr} = \frac{2\pi m_0}{qB}$ are independent of the particular lepton energy. Due to its opposite charge, an electron experiences in a B-field a Lorentz-force equal in magnitude but opposite in direction to the force experienced by a positron.

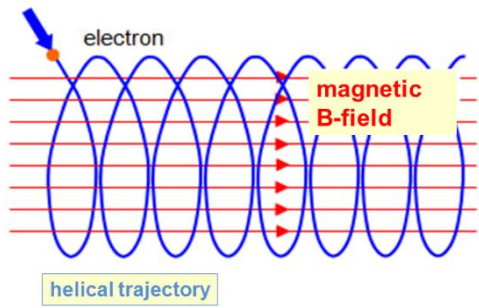

*Fig. 15. Helical trajectory of an electron around a solenoidal B field. The helix for a positron with identical initial vector momentum has opposite sense of rotation and a common tangent with the electron's helix in the starting point.*

Electrons and positrons, emitted from a single point of origin in the solenoidal B-field, execute left- and right helical motions, but around a common axis, which contains the joint point of origin and is oriented parallel to the B-field lines. However, this is not sufficient for the positive identification of the respective leptons and subsequent determination of their primordial vector momenta when both, electrons and positrons are present. In fig. 16 we compare the helices electrons and positrons emitted with momenta and azimuth angles (p-, $0^0 \leq \varphi \leq 120^0$) and (p+, $0^0 \leq \varphi \leq 120^0$), respectively, are executing with respect to the B-field in a homogeneous solenoidal B-field. For leptons emitted uniformly over azimuthal angles $\varphi$ from the target zone in a solenoidal B field, the helices of both, electrons and positrons, fill a common cylinder with a respective gyroradius $r_g$ centred on the common solenoid axis; this prohibits to differentiate electrons from positrons simply by exploiting the location on the detector.

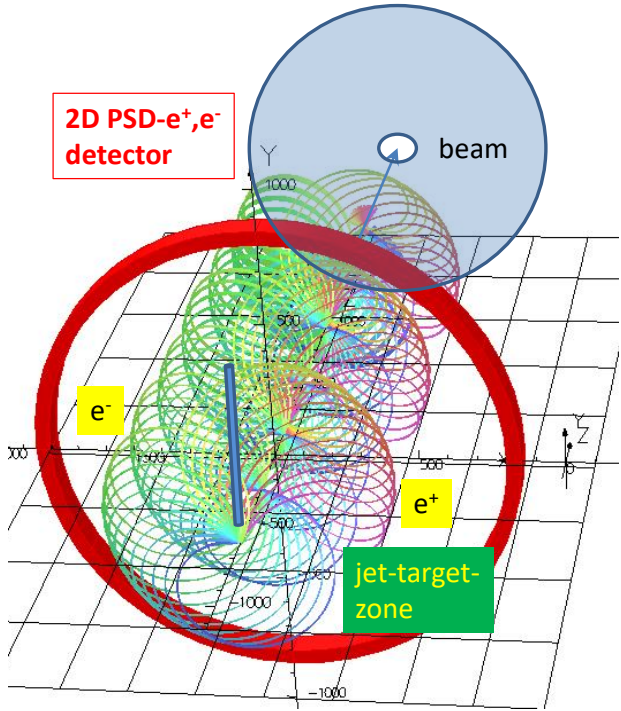

*Fig. 16. Positron trajectories with momentum p+ and electron trajectories with momentum p- launched from a common origin on the axis in the jet target zone in a solenoidal B-field oriented parallel to the beam axis. Only those leptons launched with emission directions $0^0 \leq \varphi \leq 120^0$ are displayed for reasons of visibility, for $0^0 \leq \varphi \leq 360^0$ helices would exhibit the complete overlap in space. The electrons and positrons execute left and right, respectively, helical trajectories and always intersect the axis after a completion of one full helical turn, i.e. the common revolution time T. This intersection for positrons and electrons occurs, however, at locations on the axis given by the wavelength of motion $\lambda_L$, according to the different velocity components of e+ and e- parallel to the beam axis.*

It is apparent and important to note that all trajectories of electrons and positrons always return to a point intersecting the axis containing their origin after time of flight given by T, (which in its non-relativistic expression is of course independent of the energy). The location, where the particle returns to the axis after the common revolution time T, depends on the component of lepton velocity parallel to the B field and is given by the wavelength of motion $\lambda_L$. For a given hit on the detector plane the distance to the axis even in combination with T is not sufficient to determine the vector momentum of a lepton, but most importantly, not even whether it's a positron or electron.

A solenoidal B-field has seen a powerful application in the longitudinal reaction microscope [88 and refs therein, 104], where in the analysis of the kinematically complete ionization and transfer cross sections of low energy electrons emitted in ion-atom collisions the solenoidal structure of the field permits to reconstruct the entire primordial vector momentum of electrons from their TOF and the radial displacement mapping their gyroradius in the B-field onto a 2D-PSD.

### c) Electron and Positron Trajectories in a Toroidal Field

Whereas the calculation of orbits of leptons in a pure solenoidal field is straightforward and the resulting helix a firmly established powerful tool in deriving kinematically complete ionization cross sections using reaction microscopes [88,104], the 1/R-dependence of the toroidal magnetic field

$$\vec{B} = \frac{A(i)}{R}\vec{e}_{\varphi} \tag{7}$$

(e.g. as also employed as one of the confining fields in a fusion plasma) presents a formidable obstacle for calculations of charged particle dynamics and does not allow to find an analytic solution in closed form for the Newton classical equation of motion [79,99b,120,123]. The 1/R variation (see also fig. 11) of the magnitude of the B-field across the presumed helix implies that the radius of curvature on the „inside" of the helix, i.e. closer to the center of the toroid, is smaller due to the larger B-field and is larger on the „outside" of the helix, i.e. locations further from the centre of the toroid. This entails that the orbit of the lepton, or any charged particle, will never intersect the centre line again resulting in an increasing displacement perpendicular to the B-field lines, whose direction is opposite for positive and negative charged particles, superimposed onto the helical motion.

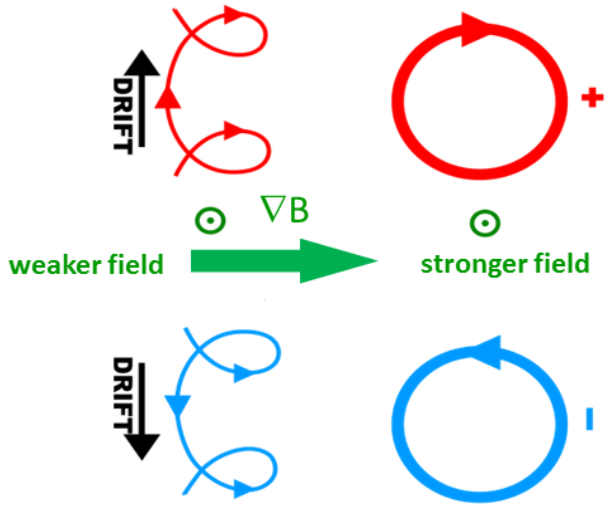

*Fig. 17. Illustration of opposite drift perpendicular to magnetic field $\vec{B}$ (with B-field direction out of the plane) and to the gradient of the magnetic field $\nabla\vec{B}$ for positively charged positrons and for negatively charged electrons*

This property of toroidal fields is the basis for its favorable disposition for lepton pair spectroscopy.

Historically Alfvén, analysing particle trajectories in magnetically stabilized plasmas, was the first to introduce an approximation [105] based on observations of the physical properties of a collisionless magnetized plasma: here the Larmor radius characterizing the gyration is much smaller than the typical variation length scale of the generating magnetic B field. This permits to decouple the fast gyromotion from the slow motion of the centre of the confining gyromotion of the electron (called the guiding centre). This technique has been used numerous times in studying the dynamics of plasma in magnetic confinement [97,110,111], but notably also for positron spectroscopy using a toroidal spectrometer, e.g. TORI, as in transient superheavy quasimolecules created in adiabatic heavy ion –atom collisions [76,77].

However, in the current context we will not impose such seriously restraining restrictions as for TORI on digression from the center line as they would confine resulting trajectories to unpractically small amplitudes of the helical motion of the lepton.

For the equations for the gyromotion and for the (slow) motion of the guiding centre across magnetic field lines in the Alfvén approximation more powerful technique have been developed in the Hamiltonian formalism [97,106-116, 123,141 ]. The key idea is to simplify the 6D (or 8D, respectively, in relativistic context) problem and to derive after some suitable canonical transformations e.g. simple Hamiltonian equations for the guiding centre and a gyrating 3D-motion of the charged particle presented as a Hamiltonian in an effective 1D potential well along the radial coordinate with slow variation along φ [96,97,110,117,118,123].

The corresponding vector potential **A** may either be approximated by analytical models or may be derived directly from given current density distributions, preferably in toroidal coordinates [100-102,141],but see also Appenix 1.

It was shown that corresponding numerical evaluations of expansion coefficients in toroidal coordinates may be performed significantly faster than standard Biot-Savart techniques [99,100-102,119-122,141] which is of relevance as guidance for further analytical analysis, e.g. for Hamilton equations of motion, is desirable.

In the present first part of our study, however, both, the B-field distribution in the toroid and solenoidal sections of the spectrometer and all corresponding lepton trajectories from production in the jet-target zone to the detector plane are calculated with OPERA-3D such that also trajectories with large gyroradii can be analysed ( for results see next section, see also Appendix 1).

## 6. Invariants and other serendipitous features of lepton trajectories in a toroidal B-field

Toroidal magnetic spectrometers permit to simultaneously investigate e[+] and e[-] continua from free-free pair production emitted in heavy ion atom collisions due to the separation of leptons of opposite charge perpendicular to the centre plane of the toroid. In the following we will illustrate the highly useful and expedient characteristics of electron and positron trajectories in toroidal B-fields and how the momentum dispersion of lepton trajectories perpendicular to the centre plane of the spectrometer permits to tune the B-field for any wanted lepton momentum window and determine the corresponding differential cross sections (DCS).

a) $\theta_{lab}^{lepton} = 0^0 \leftrightarrow p_\perp$(lab)=0 ; invariance of momentum arcs

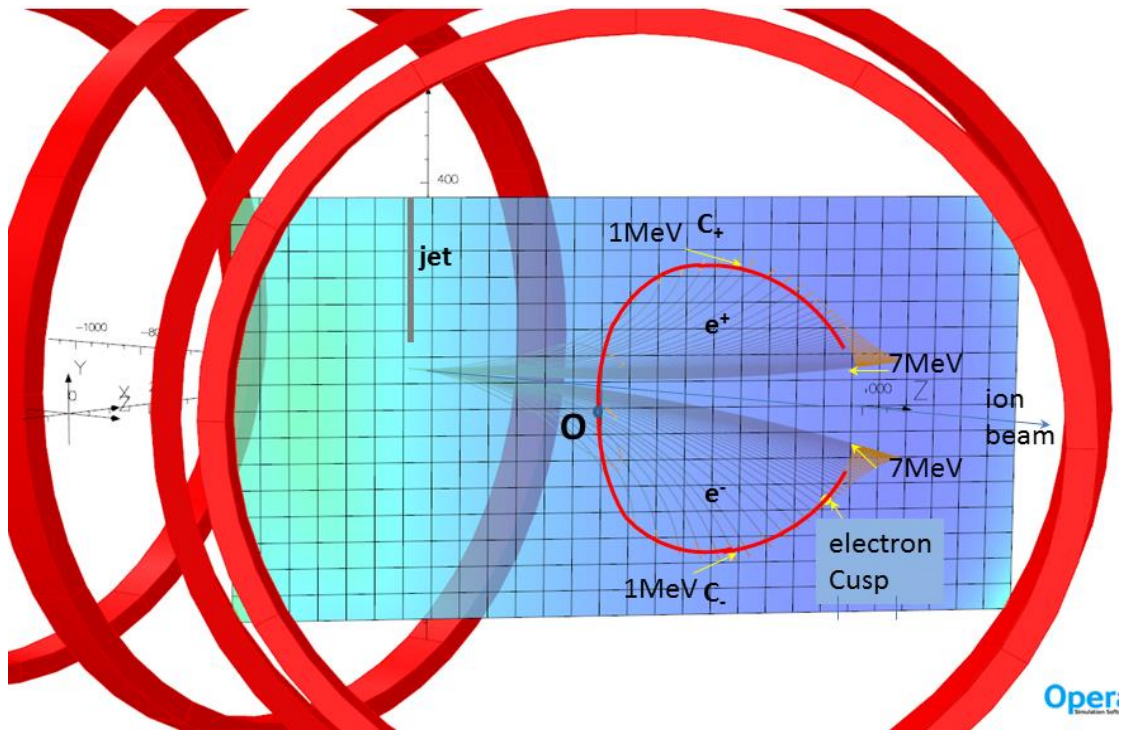

Fig 18a

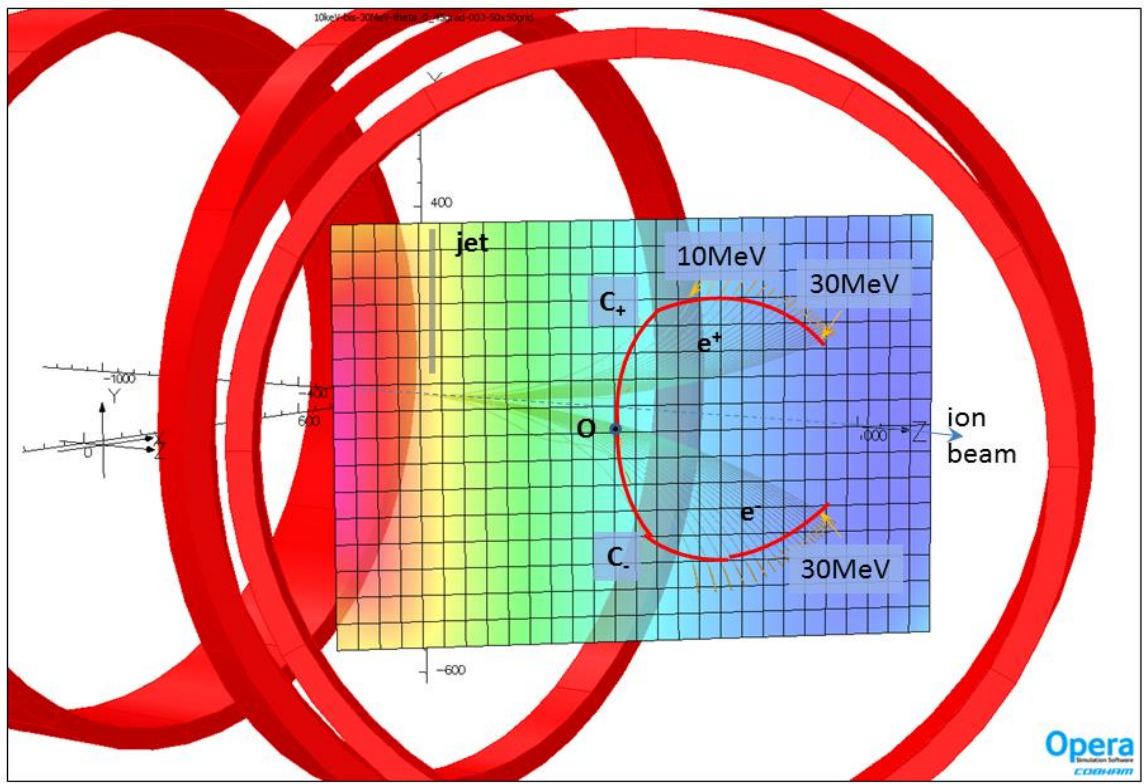

Fig 18b

*Fig. 18 a, b. Invariance of lepton momentum arcs under variation of the toroidal magnetic field. The momentum dispersion of leptons is displayed for leptons with laboratory momentum **p**=(p$_x$=0,p$_y$=0,p$_z$) in the local coordinate system with origin at the launch point in the jet target, (i.e. leptons only possessing an initial momentum*

*component in the beam direction are considered). Leptons are dispersed perpendicular to the x-z-plane = bend-plane, which is a symmetry plane of the toroidal sector. For clarity of the illustration only a few coils of the toroid including coil 7 containing the detector plane – here positioned at $\varphi=45^0$ – are shown (see fig.11 for details of coordinate frame and another option with detector plane coinciding with coil 8). For very low momenta **p** electrons and positrons with momentum p→0 will follow the B-field lines and both intersect the detector plane at the midpoint (indicated as **O**) of the coil. For a given magnetic field B, with increasing momentum of the leptons the intersection points will move away from the centre plane, the x-z- plane, and map out characteristic arcs $C_+$ for positrons and $C_-$ for electrons. For 108.7G in the upper figure the trajectories for lepton laboratory energies up to 3 MeV ($\beta\gamma=6.8$) are shown. In the lower figure one sees that geometrically identical arcs are traced out for 1087 G and lepton energies up to around 30MeV ($\beta\gamma=59.7$). For the grid shown in the respective detector planes the separation of both horizontal and vertical lines is 50 mm for the low field and for the high field case. For the low B-field the location of the electron cusp for 5.1AGeV projectile beam energy at $E_{lab}=2.8MeV$ ($\beta\gamma=6.40$) is indicated on the electron arc $C_-$; the cusp's location on the arc constitutes a tool of considerable usefulness for spectrometer calibration. B-fields given are understood as field strength on the centre ray of the toroidal bending radius $\rho$.*

In this subsection we focus on trajectories of leptons with no initial transverse momentum, $p_{transverse}=p_\perp=0$, and the formation of invariant momentum arcs in the detector plane. In fig. 18 the momentum dispersion of electrons and positrons perpendicular to the toroidal bend plane is illustrated for leptons with an initial transverse momentum $p_\perp=0$. The intersection of their trajectories with the detector plane (in figs. 18 selected to match the plane of coil 7 at $\varphi=45^0$) traces out, starting from the midpoint **O**, arcs $C_+$ and $C_-$ in the detector plane for $p_z$ increasing, for positrons and electrons, respectively; for a given B-field and the detector plane positioned at a given $\varphi$ the location of this intersection on the arcs uniquely determines the momentum of a lepton. It is important to note that for fixed $\varphi$ the arcs are invariant and their geometry does not change with a varying B-field; the geometry of the arcs only depends on the toroidal radius $\rho$ and the angle $\varphi$; we emphasize that with increasing B-field the locus of leptons of a fixed momentum will slide on the respective arc toward the midpoint **O** of the coil. By increasing/decreasing the B-field the intersection/locus of leptons of a fixed momentum may thus be moved to any desired location on the invariant arc toward the midpoint **O** of the coil or away from **O**. The actual geometric locations in the detector plane, i.e. the arcs $C_-$ and $C_+$, traced out by these electrons and positrons, respectively, starting with $p_\perp=0$ and $p_\parallel=0$ at the origin **O** in the detector plane, for all values of laboratory momentum $p_\parallel=\beta\gamma$ in the upper and lower half plane are established by optics calculations using OPERA3D [81]. This invariance of the arcs for fixed φ is an essential property of the spectrometer and will considerably facilitate the configuration of the 2D PSD detector arrays and calibration.

Apparently, the spatial resolution for low lepton momenta may be conveniently enhanced on the expense of total range of lepton momenta covered, either by properly tuning the B-field in the toroid to an adequate low field strength or by placing the detector plane at suitable larger angles φ. Low energy electrons with energies as low as 1 keV and up to around 400 keV may be conveniently studied for a magnetic field as low as 30G. The resulting increase of spatial dispersion entails, however, a reduction of the useful lepton momentum range covered in the detector plane. By moving in a 1087G field the detector plane from φ=45$^0$ to φ=73$^0$, e.g. the vertical intercept for 13 MeV leptons emitted at 0$^0$ with respect to the beam axis moves from 250mm above/below the midplane of the coils at φ=45$^0$ to a location outside the coils, theoretically near 700mm above/below the midplane, at φ=73$^0$ .

For the desired experimental momentum calibration of the electron-arc $C_-$ one has a supremely powerful and very practical tool: the prominent electron cusp [124, and references therein], which contains electrons emitted with very low $E_{kin}$ in the emitter frame and under near 0$^0$ with respect to the projectile direction and thus with velocity $v_e \cong v_{par} \cong v_{proj}$ (i.e. $v_\perp$=0). The location of the cusp on the arc $C_-$ for varying B-field or for different projectile energies will map out and calibrate locations corresponding to cusp lepton momentum $(\beta\gamma)_{lepton}=(\beta\gamma)_{proj}$ on this arc $C_-$. The symmetry of trajectories for electrons and positrons in the toroidal magnetic field facilitates construction of the homologue arc $C_+$ for positrons as mirror image of $C_-$ with respect to the xz bend plane of the toroidal configuration.

b) $\theta_{lab}^{lepton} \geq 0^0 \leftrightarrow p_\perp(lab) \geq 0$;

In the following we also consider leptons with initial transverse laboratory momentum $p_\perp$>0 and show how the imaging properties of a toroidal field in combination with a set of 2D position- and energy sensitive detectors permit to retrieve all laboratory momentum components, $p_{transverse}$ and $p_{parallel}$, of both leptons and thus obtain then the desired angular emission characteristic of the leptons in the emitter frame. In the remainder of this section we restrict discussions to laboratory emission angles $\theta_{lab}^{lepton}$ around 20$^0$ and below, which in view of fig 9 amounts to a tolerable restriction of the emitter frame phase space of the leptons.

Some characteristic features of lepton trajectories described in the previous section for solenoidal fields (see also fig. 16) will be approximately conserved and will persist with some adequate modifications also in toroidal fields [see ref. 82 for case of low energy electrons in a toroidal reaction microscope]: trajectories for leptons of momenta $p=\beta\gamma/n$, $n\geq 1$ and not too large forward emission angles mostly not extending beyond $20^0$ with respect to the beam axis, i.e. $p_\perp(lab)\geq 0$, will intersect at or very near a focus $X_n$ when their time of flight (TOF) $T=S_j/\beta_{par}c$ ($S_j$ =arc travelled by lepton with perpendicular momentum $p_{perp}=0$ to $X_n$ in detector plane j) from the origin at the jet target to the focus $X_n$ (determined by their longitudinal momentum $p_{par}$) is a multiple of the inverse of the effective cyclotron frequency $f= qB/2\pi\gamma m_0$ for the B field given:

$$T = \frac{S}{\beta_{par}c} = nf^{-1} = n\left(\frac{qB}{2\pi\gamma m_0}\right)^{-1} = n\frac{2\pi\gamma m_0}{qB} \qquad (9)$$

It is then straightforward - by tuning the B-field - to position the anticipated focus $X_n$ for a well-defined lepton momentum $p_n$ on the respective arc in a chosen detector plane (recall that the geometry of an arc is only depending on the detector plane location and the radius of curvature $\rho$ of the toroidal field, but is invariant under variation of the B-field).

In fig 19 we show trajectories in a 108.7G field for leptons with two different momenta in the laboratory, $\beta\gamma=1.2/n$, and $1\leq n\leq 2$, respectively, for laboratory emission angles up to $\theta_L=20^0$ which intersect for all azimuthal angles in four distinct foci on the arcs $C_-$ and $C_+$.

The usefulness of the relation given in eq. 9 is, of course, contingent on the assumption that only trajectories in the confidence volume of the magnetic field B is considered (recall that following eq. 7 the B-field inside the toroid strictly varies with 1/R). This implies, for the foci to be useful (i.e. the foci exhibiting a negligible astigmatism and thus reasonably small extension) a restriction to excursions perpendicular to the centre of curvature $\rho$ of the toroid (see fig 11) not extending into the fringes of the coils, i.e. a moderate gyro-radius $r_g$ of the lepton, is imposed. This assumption is met in the example illustrated in fig 20.

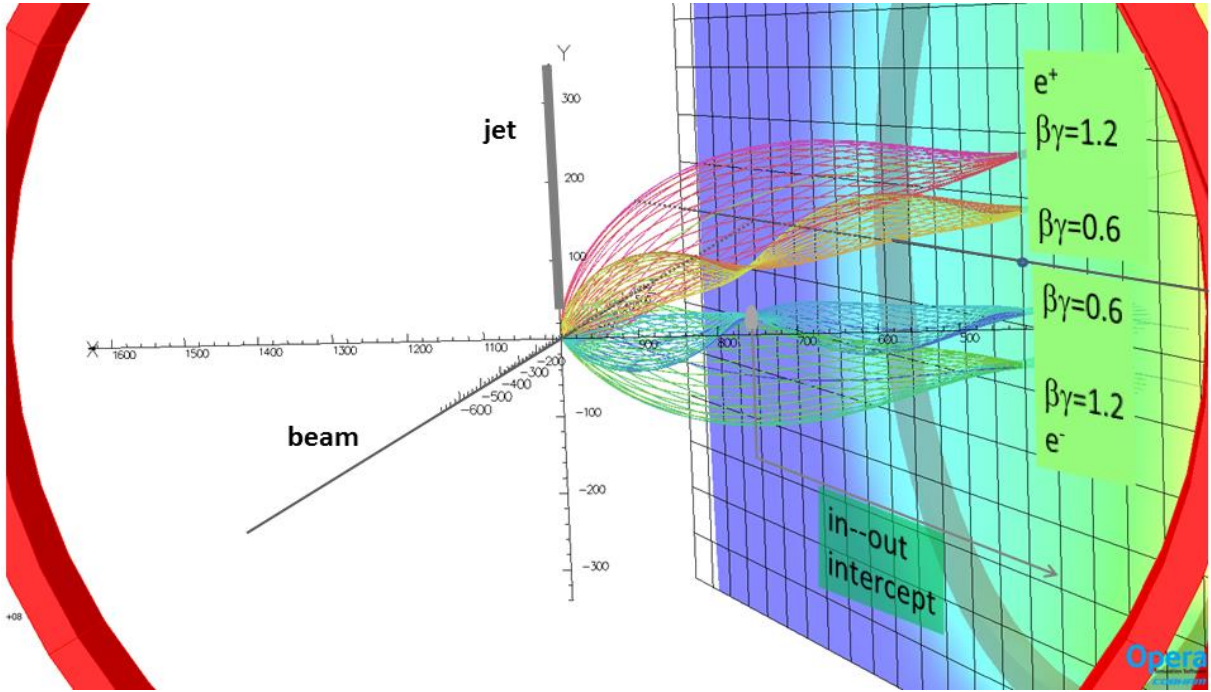

Fig. 19. Electrons and positrons launched at the supersonic jet target (Cartesian coordinates (1000, 0, -50), see fig. 11) with polar angles $\theta_L=20^0$ and azimuthal angles $0^0 \leq \Theta \leq 360^0$ with respect to the projectile axis = local z-axis at the jet. In the given configuration with B=108.7 G and the detector plane located at $66^0$ all lepton trajectories with $\beta\gamma=1.2/n$, n=1, 2, are intersecting at well distinguishable locations $X_n$ in the detector plane, where an array of 2D PSDs may be appropriately configured, centred on the arcs $C_+$ and $C_-$. The lepton kinetic energies corresponding to the trajectories given here are 85 keV ($\beta\gamma=1.2/2$) and 287 keV ($\beta\gamma=1.2/1$), respectively. The detector plane has grid-tiles of 100mm x 100mm. Note, that trajectories for different azimuth $\psi$ (in coordinates with origin at jet) exhibit the completion of one full circle upon intersection at the first focus following eq. 9 for leptons with 287 keV. For electrons with $\beta\gamma=1.2/2$ leptons have completed two full circle upon intersection at the detector plane; the option of an intercept of electrons at a suitable intermediate focus by a small disc or even a 2D PSD is illustrated for the lower energy electrons with $E_{lab}$=85keV, i.e. $\beta\gamma=0.6$.

For any B-field selected these foci $X_n$, $n \geq 1$, in the detector plane are located along the arcs $C_+$ and $C_-$ according to their momentum $p_n=(\beta\gamma)/n$ shown in fig 18. Lepton trajectories for leptons with $p_n=(\beta\gamma)/n$, n>1, contain (n-1) intermediate foci before hitting the nth focus in the detector plane at $X_n$. With increasing n, i.e. decreasing momenta, the spatial separation of foci $X_n$ located on the arcs $C_+$ and $C_-$ decreases rapidly and foci $X_n$ converge towards **O**; this implies that the entire backward emission of fast leptons (as seen in the projectile frame, see fig. 9) out of the fast projectile will thus virtually collapse to one-dimensional arc sectors in the detector plane either side of **O**.

The corresponding single differential cross section $\frac{d\sigma}{dp}$ for lepton production mapped onto a 2D PSD may thus be conveniently recovered. It follows that the coincident DDCSs $\frac{d^2\sigma}{dpd\Omega}$ for free-free pair production where both leptons are emitted into the backward direction (in the emitter frame) is similarly acquired in a straightforward fashion via coincident detection of electrons and positrons in the corresponding detectors covering the arcs close to **O**. (In passing we note that from fig 9a one sees that under $\theta_{lab}$ in the vicinity of $0^0$ for decreasing laboratory lepton energy/momentum only leptons with increasing emitter frame energy are contributing).

For free-free lepton pairs with backward-forward emission pattern in the emitter frame correspondingly the coincident laboratory frame DDCS are derived from coincidences between a lepton on the arc section close to **O** and the coincident second lepton of the pair at individual foci $X_n$, $n \geq 1$.

Leptons with momenta $p > p_1$ will not exhibit an interception focus independent of perpendicular momenta in the detector plane (see also fig 26); this enables direct determination of full laboratory DCS for $p > p_1$.

The apparent near convergence onto a focus $X_n$ after the time of flight TOF $= n\, f^{-1}$, with f the cyclotron frequency, is obviously facilitated and its usefulness augmented by the considerable kinematic restriction to small laboratory angles $\theta_{lab}$ arising at large collision energies as provided by the HESR, for which $p_{par}$ and thus the TOF only exhibit a weak dependence upon $\theta_{lab}$ (see also fig. 9). However, for large laboratory transverse momenta, i.e. for gyro-radii $r_g = \frac{m\gamma\beta_\perp c}{|e|B}$ corresponding to lepton trajectories sampling regions with B-fields noticeably larger (at $R < R_0$) and smaller (at $R > R_0$) than the B-field at the centre radius $R_0$ of the toroid, the inverse cyclotron frequency varies sufficiently as to generate azimuth-dependent interceptions [82] instead of single nearly azimuth-independent foci $X_n$ as seen for the small gyro-radii $r_g$.

This salient feature of lepton trajectories in a toroidal magnetic field – appearance of periodic intermediate foci along trajectories to the detector plane for not too large gyroradii - may also be employed to intercept electrons and positrons of any desired momentum $p < p_1$ at suitable locations before the detector plane with high efficiency by small diameter detectors or beam stops, with minor effects on leptons of other momenta due to the strong spatial dispersion. This permits to introduce for particular combinations of lepton energies highly efficient multiple detector planes (see e.g. single intercept in fig. 19), as also the location of the intermediate foci are space fixed and do not depend on the magnetic field B ( for details see section d).

Figures 20 show for a high B-field of 1087 G the intersection in the detector plane of lepton trajectories over a wide range of laboratory energies and two sets of emission angles. This corresponds to covering emitter frame energies up to 500keV and 1MeV (which is the mean lepton energy $E_{av}$ predicted by Scheid et al. for a similar collision velocity) and even above $E_{av}$ for the highest laboratory energies (see also figs. 9 and 10).

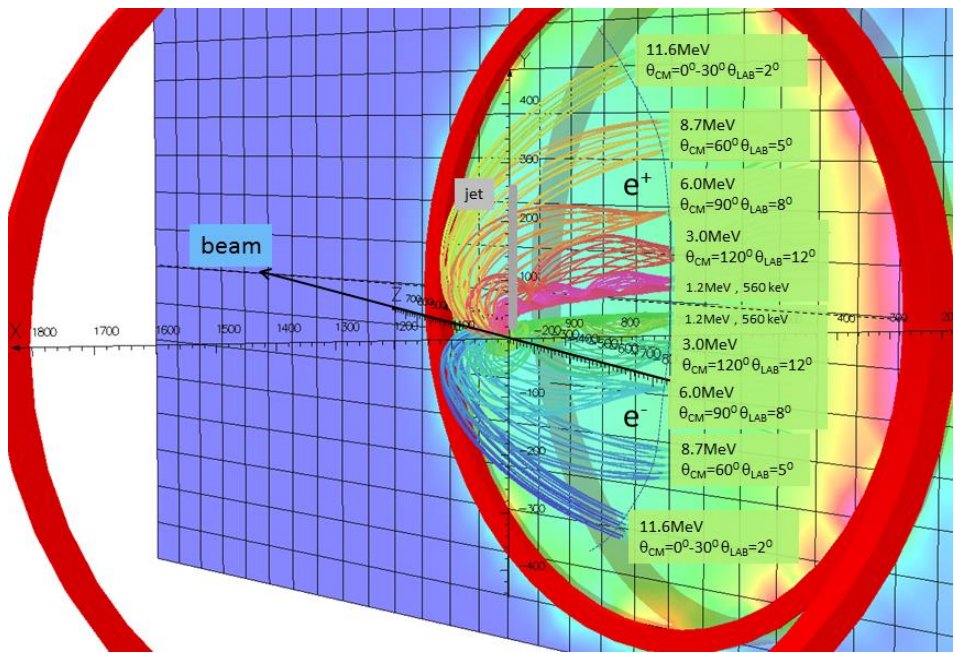

*Fig. 20. Lepton trajectories in the toroidal B-field =1087G. For the laboratory lepton energies selected here the range of projectile frame angles and laboratory frame angles are given. The arc $C_+$ and $C_-$ (see fig. 18) along which leptons are dispersed are indicated as dotted lines. The tile dimensions are 100mm x100mm. For illustration only lepton trajectories with $p_n=\beta\gamma/n$ for $\beta\gamma=12.7$ and [n=1, 4] are displayed besides two cases with higher lepton momenta. Intercepts in the detector plane for $p_n$ with $n \geq 2$ then easily produce laboratory differential cross sections for backwards emission (as seen in the emitter frame, and summed over a wide range of emitter frame energies). The spatial overlap of lepton intercepts for the larger $\theta_{lab}$ will not confuse the correct identification of lepton energy as the 2D-PSD detector may identify the lepton energy due to energy resolution down to $\Delta E/E \approx 1\%$.*

For completeness we add that the entire very-low energy electron emission out of the projectile ($E_{kin} \leq 1$-2keV, again as seen in the projectile frame, in contrast to high $E_{kin}$ around 500keV to 2000keV as in fig. 9) is well known to be entirely contained in the electron cusp [124 and refs. therein], which for 5.1AGeV specific projectile energy $E_{beam}$ is found at an electron laboratory kinetic energy of $E_{cusp}$ =$E_{beam}$/1822.8878= 2.8MeV at $\theta_{lab}\approx0^0$ (see also the cusp in fig 9 as the point of convergence in the laboratory for emitter frame energy $E_{emitter} \rightarrow0$). The low energy electrons in the cusp have been shown to mostly arise over a range of close to more peripheral collisions (i.e. with larger internuclear separations) from radiative and non-radiative electron capture into the projectile continuum, RECC and ECC, respectively and electron loss to continuum, ELC, for non-bare projectiles [124].

 In the domain of very low energy (emitter frame) leptons besides above mentioned cusp electrons also positrons may appear, originating in free-free as well as in bound-free pair creation with very-low kinetic energy in the emitter frame; they will then appear in a corresponding mirror location on the $C_+$ arc. Positrons from bound-free pair creation (eq. 2a) will be unambiguously identified via coincidences with a projectile with an outgoing charge reduced by one unit from the incoming charge state and thus are clearly distinguished from free-free lepton pair creation (eq. 1a) and their identification via electron-positron coincidences.

Any electron emission from direct target ionization is, of course, not subject to kinematic restrictions concerning transverse momenta discussed above; the corresponding low energy electron distribution will be rather isotropic and their energy distribution is a continuum with a near exponential fall-off above zero.

c) General momentum/energy calibration of lepton trajectory intersections in the detector plane, i.e. including $p_\perp$(lab)$\geq$0.

In the general case, i.e. the absence of any kinematic restrictions for coincident lepton pairs (as e.g. back to back emission in the emitter frame), a unique attribution of the primordial vector momentum of the lepton in the emitter frame (the emitter frame characterized by $\beta_0$ and $\gamma_0$ , a lepton in the emitter frame by p', β', γ')  using

$$(p'c)^2 = (pc)^2 \sin^2 \theta_{lab} + \gamma_0^2\Big(\text{pc}\cos\theta_{lab} -\beta_0\sqrt{(pc)^2 + m^2c^4}\Big)^2 \quad (10)$$

can only be provided when both, the lepton's laboratory momentum p or energy $E_{lab}$ and the laboratory observation angle $\theta_{lab}$, are simultaneously determined, as follows from fig. 9. The corresponding equation relating laboratory frame and emitter frame angles is

$$cos\theta' = \frac{-\frac{\beta_0}{\beta'}\gamma_0^2 sin^2\theta_{lab} \pm \sqrt{\gamma_0^2(1-(\frac{\beta_0}{\beta'})^2)sin^2\theta_{lab}+cos^2\theta_{lab}}\ cos\theta_{lab}}{\gamma_0^2 sin^2\theta_{lab}+cos^2\theta_{lab}} \qquad (11)$$

where $\beta'$ corresponds to the momentum of the lepton in the moving frame.

Instead of directly measuring $\Theta_{lab}$ of a lepton, however, it is far more convenient to experimentally determine its gyro-radius $r_g$ in the toroidal B-field and subsequently calculate its laboratory transverse momentum. The relation between the true gyroradius $r_g$ and the effective $r_{g\text{-eff}}$ obtained from the location of the intercept in the detector plane is, however, best established in the spectrometer calibration using conversion electron lines, because only for locations outside the foci $X_n$ on the arcs $C_+$ and $C_-$ in the detector plane one may eventually obtain $r_g$ and the transverse momentum and its azimuth from $r_{g\text{-eff}}$ (see below).

The first step of the experimental momentum/energy calibration of the arcs $C_+$ and $C_-$ uses the narrow electron cusp with $(\beta\gamma)_{Cusp} = (\beta\gamma)_{beam}$ and energy $E_{cusp}$ =$E_{beam}$/1822.8878 correspondingly ; ( for cusp electrons $\theta_{lab}$=$0^0 \leftrightarrow p_\perp$(lab)=0 to a very good approximation) at varying beam energies (as indicated in the previous sections). Additionally conversion electron lines from radioactive sources equipped with an aperture for $\theta_{lab}$=$0^0$ will be employed. We may thus attribute for any suitable combination of B-field and projectile energy to every point $X_E$ on the arc $C_-$ an electron energy E and electron momentum p, as illustrated in figs 18.

For electron transverse momenta $p_\perp$(lab)$\geq$0 intersections on the detector plane off the arc $C_-$ appear; here suitable conversion electron lines from radioactive sources (now optionally equipped with selected angle defining apertures for distinct $\theta_{lab}$>$0^0$ resulting in well-defined $p_\perp$(lab)$\geq$0 for every angle $\theta_{lab}$ ) are employed  (see figs. 21, 22).

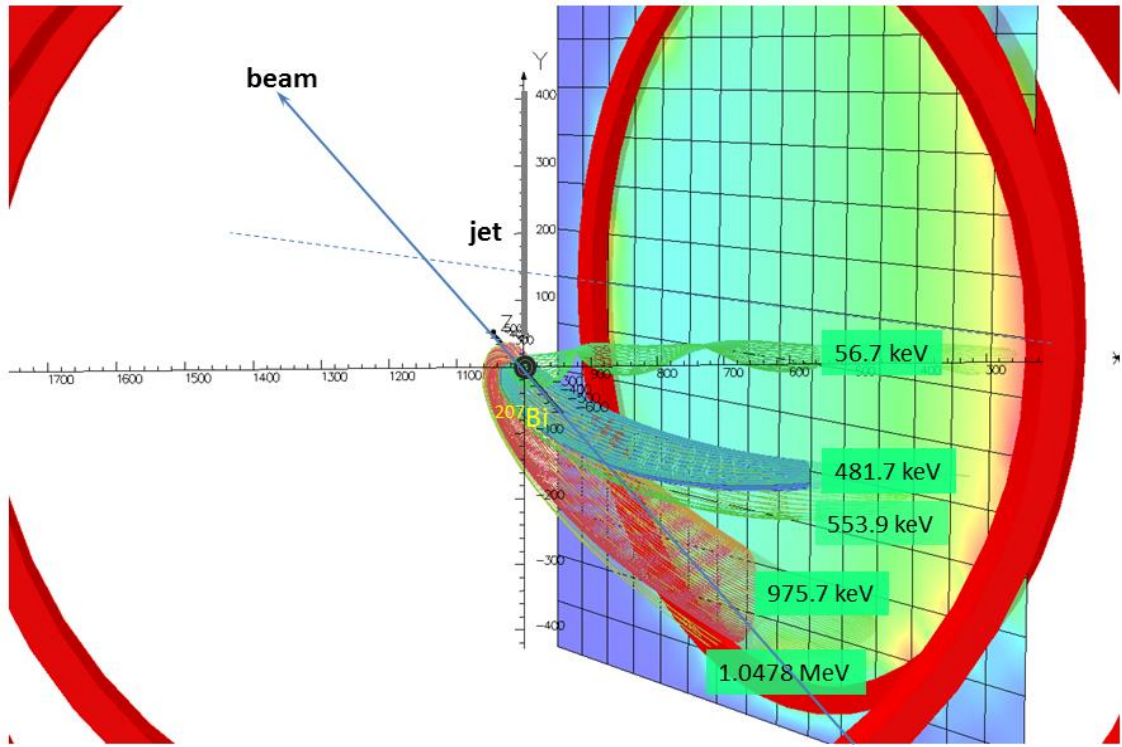

*Fig. 21. Momentum calibration with a $^{207}$Bi source in 108.7G toroidal magnetic field. Energy/momentum calibration of the toroidal magnetic spectrometer using Pb KLL Auger and conversion electron lines emitted by a $^{207}$Bi source and detected at $66^0$ for a B-field setting of 108.7G. Besides the Pb KLL Auger group centred at 56.7 keV we show in this illustration only the strongest components of the conversion electron line groups; besides the strong 481.7 KeV line the 553.9 keV line with $\approx 25\%$ intensity of the former and the 975.7 keV line with the 1047.8 keV line which also has around a quarter the intensity of the former. The emitter source situated beneath the jet in the target zone is furnished with an angle defining aperture, which can be suitably adjusted to angles between 0 and 20 degree, here an angle of $5^0$ is used in the trajectory calculations. The use of various angles on the source serves to calibrate the effective gyro-radius seen in the detector plane. The combination of $\approx 1\%$ energy resolution and 2D position sensitive detection of the electron accomplishes in an experiment the unambiguous identification where electrons of different energies may hit the same location on the detector.*

For the general gyro-radius/momentum calibration, i.e. intercepts in the detector plane off the arc C-, it is convenient to use angle defining apertures, e.g. $0^0$, $5^0$ and $15^0$ with respect to the beam axis, for a $^{207}$Bi conversion electron source over a range of different magnetic B-fields in the toroid. This permits to calibrate the apparent gyroradius $r_{g-eff}$ for the respective transverse momentum component (and the corresponding true gyro-radius $r_g$) selected by the angle setting in the aperture, e.g. of the prevailing 975 keV conversion electron line over a range of magnetic fields. In fig. 22 we show the mapping of the 975keV conversion electron line for a range of B-fields.

For 165 G we illustrate in fig. 22 how using apertures for two different angles the procedure for momentum calibration of the locations of the intercept in the detector plane is accomplished. It can also be nicely seen that for 234G all electrons from the 975.7 keV line emerging in the forward direction with θ≤5⁰ exhibit a focus after 66⁰, i.e. at coil 8 of the toroid, in one location on the arc C⸱.

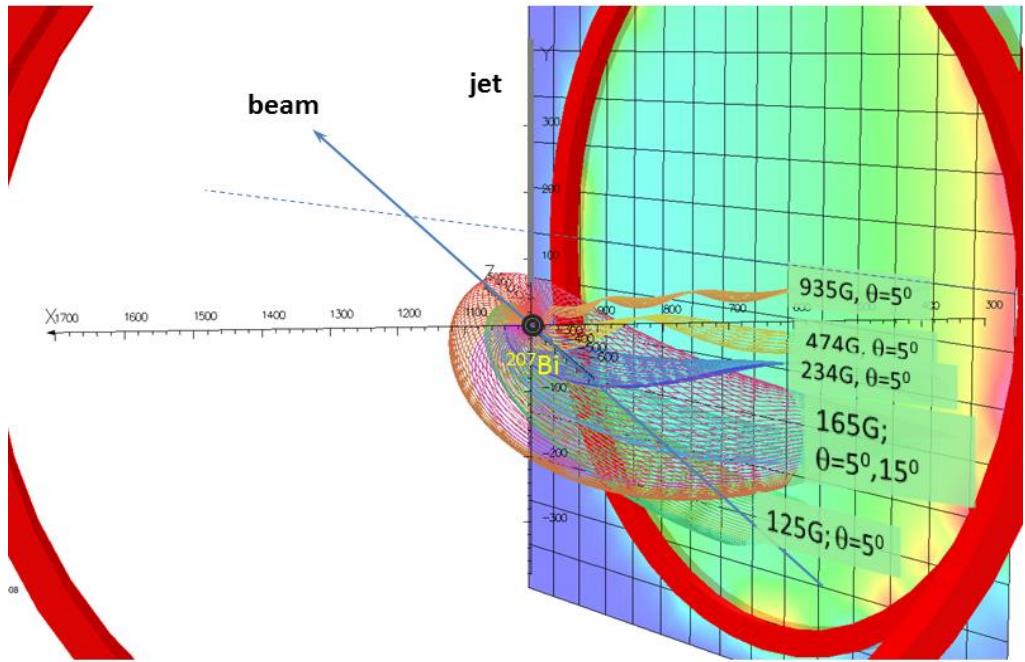

*Fig. 22. Mapping of the 975.7 keV conversion electron line from $^{207}$Bi onto the detector plane at 66⁰ and toroidal magnetic B-field for fields ranging from 125 G top 935 G. The $^{207}$Bi source located in the target zone beneath the supersonic jet target is in this OPERA simulation equipped with angle defining apertures to better illustrate the mapping used for calibration. It is apparent that for the electron energy selected, 975 keV, the B-field may either be tuned to focus all electrons onto one focal spot in the detector plane, as here shown for 234G, or to any lower value, e.g. here 165 G, where the effective gyro-radius on the detector plane can be related to the transverse electron momentum via the known laboratory emission angle in the calibration source.*

On the path to obtain differential cross sections in the emitter frame, it is instructive to look at leptons emitted with a fixed emitter frame energy, but over a wide range of emitter frame angles; they will appear over a wide range of energies and angles in the laboratory frame; this results in very useful allowed "bands" on either side of the arcs as is illustrated in fig. 23a,b.

Figs. 23a,b exhibit for 2 magnetic field settings, 272G and 1087G, for an emitter frame kinetic energy of 0.2MeV (βγ=0.96) and 1MeV (βγ=2.8), and 0.1, 0.2, 0.5, 1 and 2 MeV, respectively, for emitted leptons the regions covered in the detector plane around the arcs C$_+$ and C$_-$ for leptons of the corresponding emitter frame energies. It is apparent that 2D PSD detectors do not have to cover the entire cross section of the coil in the detector plane but that the leptons are restricted to a reasonably confined area either side of the arcs in the detector plane, e.g. as shown at φ=45$^0$ .

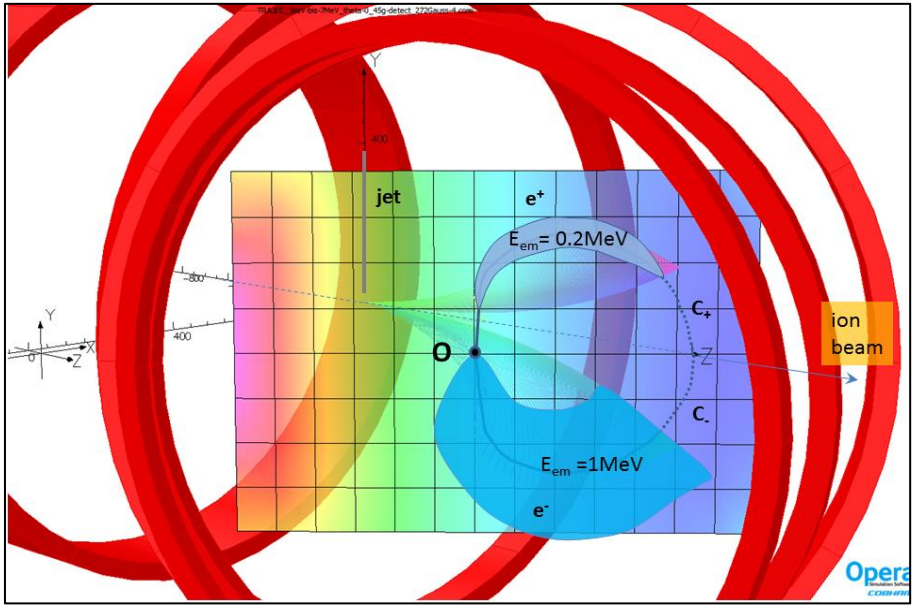

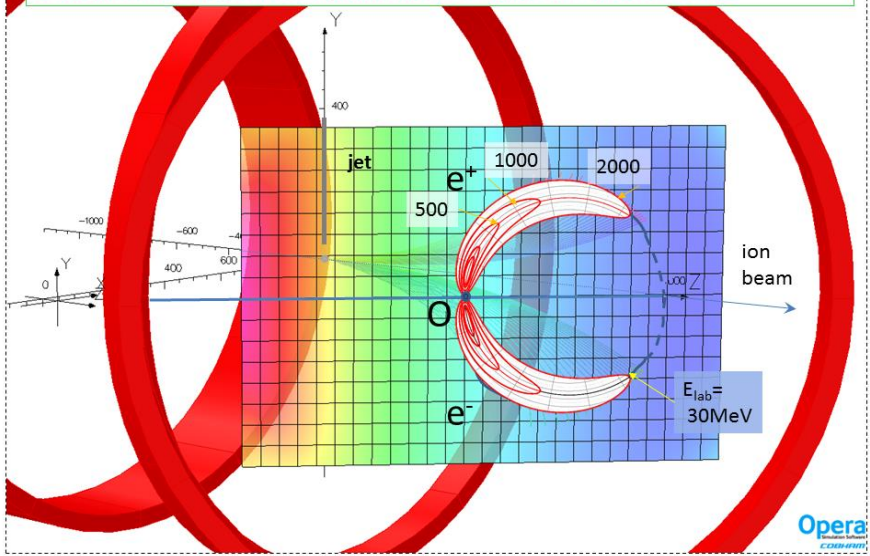

*Fig. 23a,b. Kinematically allowed regions in the detector plane for leptons emitted by 5.1AGeV projectiles for two B-field settings of the toroidal magnetic spectrometer with the detector plane at φ=45$^0$. The top figure displays for B=272G the regions adjacent to the arcs C$_+$ and C$_-$ where leptons with 0.2MeV and 1 MeV kinetic energy in the emitter frame can be detected; in the lepton trajectory calculation for the low B field only laboratory energies up to 7 MeV were used, for E$_{Emitter}$=1MeV this corresponds to*

$p_\perp$(lab)≥0. *In the lower figure the corresponding regions for leptons with 0.1, 0.2, 0.5, 1 and 2MeV are displayed for a magnetic field of 1087 G.*

The areas either side of the arcs exhibit how possible locations of the intersections of lepton trajectories with the detector plane are confined. In these "allowed regions" now a fine structure evolves superimposed on these areas in accordance with the foci in the mapping of lepton trajectories in the toroidal B-field, following eq. 9 (also figs. 19-22). This turns out to be extremely useful for obtaining differential cross sections and will be discussed in the following.

For electrons detected at a location P in the lower half plane on the 2D PSD detector its total kinetic energy $E_l$ measured attributes uniquely a corresponding location $X_E$ on the arc C$_-$, respectively, of an electron with kinetic energy $E_l$, but vanishing transverse momentum. The distance $\overline{PX_E}$ of the actual location P of the hit to the location $X_E$ determines the apparent gyro-radius $r_{g-eff} = \overline{PX_E}$. For leptons with momenta approximately fulfilling eq. 9, i.e. the intersection of their trajectories in the detector plane converging on a focus, obviously the distance $\overline{PX_E}$ will be vanishingly small. For other locations $\overline{PX_E}$ will underestimate the true gyro-radius $r_g$ of a lepton by a location dependent factor (see fig 24). For this reason for a given magnetic field B the true gyro-radius $r_g$ has to be derived always from calibrations with conversion electron lines from the measured apparent gyro-radius $r_{g-eff}$. The momentum calibration found for electrons can then be applied in analogy for positions of positrons in the upper half plane.

The true gyro-radius $r_g$=$(\beta\gamma)_\perp m_0 c/qB$ then determines $p_\perp$=$(\beta\gamma)_\perp$ of the lepton, as illustrated in fig. 24. $E_l$ and $p_\perp$ now uniquely determine for the detected lepton the emitter frame energy (see fig. 10b) and also its emission direction in the emitter frame (fig.9).

Due to the invariance of arcs C$_+$ and C$_-$ at a given toroidal angle $\varphi$ , all leptons with $p_{perp}$ >0 will intercept the detector plane in well-defined areas on either side of the arcs C$_+$ and C$_-$ as illustrated in fig 23. This will considerably facilitate design of the appropriate 2D PSDs detector frames for lepton detection.

In figs 24 we illustrate the relation between apparent gyro-radius and the transverse laboratory lepton momentum for two B-field settings, B=1087G and B=272 G.

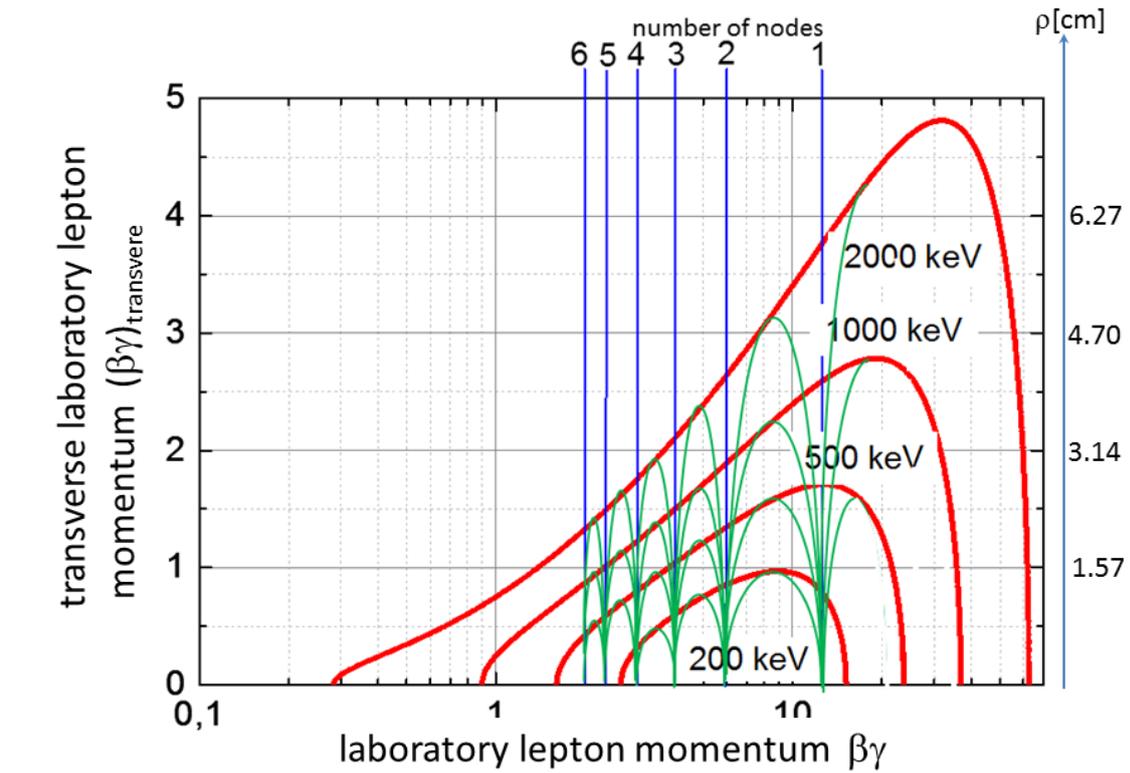

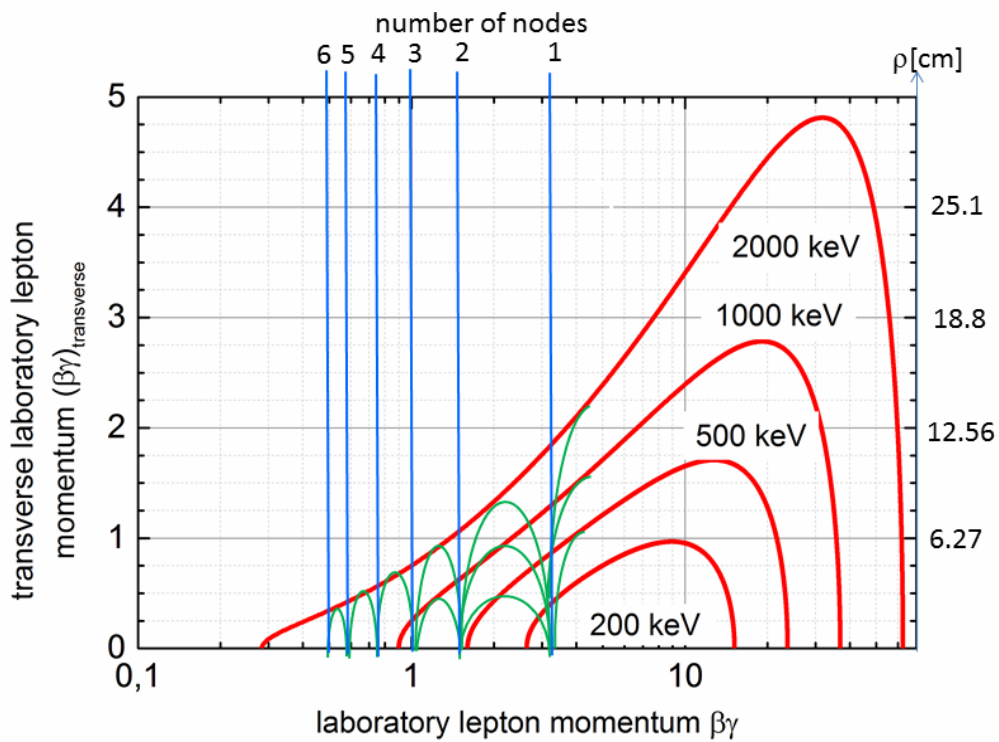

Fig. 24 a b. Transverse lepton momentum $p_{transverse}$ and true and apparent gyro-radii $\rho$ vs. laboratory momentum $\beta\gamma$ for four emitter frame lepton energies 200, 500, 1000 and 2000 keV at emitter specific energy of 5.1 AGeV for two magnetic fields, 1087G and 272 G. The red curves represent the relation between a lepton's transverse momentum *and total laboratory momentum for the four selected emitter frame lepton energies. The vertical blue lines indicate those laboratory lepton momenta ($\beta\gamma$), for which according to eq. 9 at the two magnetic fields B=272G and B=1087G foci are generated at the detector plane, i.e. all trajectories of this momentum will intercept at one locus in the detector plane (here at $\varphi=66^0$) with an apparent gyro-radius=0, thus not permitting determination of $\theta_{lab}$ or transverse momenta; numbers above the blue lines give the number of nodes completed when the trajectory crosses the detector plane: n=1 means first focus in detector plane, n=2 means one intermediate focus: (n-1) gives the number of intermediate foci (for momentum $p_n$) before hitting the detector plane (see also fig. 19). The ordinate on the right in the two figures gives the particular gyro-radius for the leptons in the two respective B-fields using $\rho[m]= 1.7045\ 10^{-3}\ (\beta\gamma)_{perp} /B[T]$, e.g. $\rho[cm]=6.27\ (\beta\gamma)_{perp}$ for B=272G. Vertical lines for lepton momenta $p_n$, n>6 will appear with increasing density and have thus been omitted for simplicity of the figure. Analogousiy for a field B=272G the group of lines is shifted to lower momenta on the abscissa, the first node appears at $\beta\gamma=3.15$ and all subsequent nodes correspondingly at lower momenta $\beta\gamma$/n, n$\geq$2. Red lines represent the true gyro-radii and the green lines between the nodes represent the apparent gyro-radii $r_{g\text{-eff}}$ to be determined by calibrations. Importantly, the geometric location of the nodes in the detector plane is invariant under change of B-field ( see below).*

In figs 24 we illustrate for a collision energy of 5.1AGeV and four emitter frame lepton energies between 0.2 and 2MeV the laboratory transverse momentum of the lepton as function of laboratory momentum $\beta\gamma$ and corresponding gyro-radii in the two B-fields which is in a satisfactory approximation calculated using here the equation for a homogenous B-field: the true gyro-radius $r_g$ for a lepton with momentum perpendicular to the field follows from $r_g[m] = 1.7045\ 10^{-3}(\beta\gamma)_{transv}/B[T]$. The toroidal nature of the field, however, produces intermediate foci along the path to the detector plane as outlined above. For a 1087G toroidal B-field and the detector plane at $66^0$ these foci appear at momenta indicated by six vertical lines in fig. 24 (similarly for the 272G field the lines appear at corresponding lower momenta). At these momenta leptons will intersect almost independent of their transverse momentum $p_{transverse}$ at one focal point on the detector plane. For these momenta the transverse momentum is integrated over and no determination of $p_{transverse}$ is possible, as the apparent gyroradius $r_{g\text{-eff}}$ is 0. For momenta between the momenta $p_{n-1}$, $p_n$ of the foci a determination of true $p_{transverse}$ and true gyroradii via measured apparent gyro-radii is possible as outlined before; this is indicated by the green arcs centred on the foci at the blue vertical lines in figs. 24a,b For lepton momenta situated approximately between two foci a measured apparent gyro-radius (to be imagined as situated on a virtual green arc) needs to be corrected accordingly using the calibration to derive a true gyro-radius $r_g$ and the correct transverse momentum, on the corresponding red line; - this now allows to determine the emitter frame energy and emission direction of the lepton.

In the calibration with a conversion electron source the angle defining aperture on the source determines for every electron line (e.g. Pb KLL-Auger and 500keV or 1000 keV conversion electron lines from $^{207}$Bi) of momentum $(\beta\gamma)$ the transverse momentum component $(\beta\gamma)_{transverse,}$ and thus a corresponding "red" point in a diagram analogue to those in figs 24 for a suitably chosen toroidal B field. The actually measured "green" point in the calibration determines for the momentum $(\beta\gamma)$ on the abscissa an apparent gyro-radius $r_{g\text{-eff}}$ and thus a correction factor. This is to be executed for various angles of the source and all suitable conversion lines, i.e. lepton energies on the abscissa. Thus correction factors are obtained to derive for measured hits in the detector plane the true $(\beta\gamma)_{transverse}$ at this abscissa.

For lower B-fields leptons with correspondingly lower momenta will be focussed, but onto exactly the same geometric locations $X_n$, $n \geq 1$ (see below). (This is of considerable usefulness for designing the 2D-PSD detector configuration and the calibration procedures, as well as for the subsequent derivation of emitter frame DCS).

A full identification of a lepton's primordial vector momentum in the emitter frame just requires the determination of the lepton's laboratory energy E and of the hit location P in the detector plane in order to determine $p_{transverse}$. This is the basis for the next step, determining the differential production cross sections in the emitter frame.

    d)  geometric invariance of location of foci on arcs $C_+$ and $C_-$

As outlined already above, for a given B-field in the toroid the collision velocity and emitter frame lepton energy and -emission direction ( i.e. lepton vector momentum) determine the kinematically allowed regions adjacent to the arcs $C_+$ and $C_-$; they are highlighted for selected emitter frame lepton energies in figs 23a,b. These regions will, however, not be uniformly illuminated by the leptons due to the characteristics of lepton trajectories in a magnetic toroidal field.

The near helical character of the lepton's orbits in the toroidal field and the axial intercepts at well determined momenta $p_n = \beta\gamma/n$, $n \geq 1$, resulting in a series of foci $X_n$ on the arcs $C_+$ and $C_-$, will modify the mapping of each lepton's phase space onto the 2D-PSDs, as illustrated in figs. 24 and 25. When comparing the location of the foci on the arcs $C_+$ and $C_-$ for 3 different B-fields, we recognize some highly significant features:

i)    as exhibited above( fig 18a,b) for leptons with $p_{perp}=0$ the geometry of the arcs $C_+$ and $C_-$ for a given toroidal radius $\rho$ and angle $\varphi$ is invariant under the change of the toroidal B-field,

ii)   with increasing B-field leptons of correspondingly increased momenta $\beta\gamma$ are focussed,

iii)  <u>and most importantly</u>, also the geometric locations of the foci $X_n$ on the arcs $C_+$ and $C_-$ are invariant under a change of toroidal B-field. The momentum of these leptons at location $X_i$ increases proportional to the B-field.

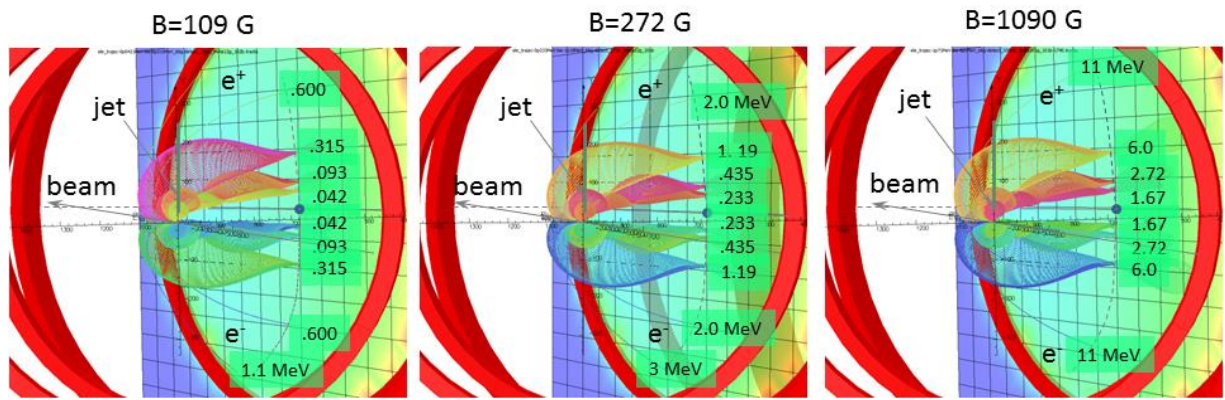

*Fig. 25. Invariance of geometric location of foci under variation of the toroidal magnetic field, illustrated for foci $X_i$, $1 \leq i \leq 3$, on the arc $C_+$ and $C_-$ at fixed position of the detector plane at $\varphi=66^0$. For three B-fields the lepton energies at these invariant locations, which exhibit a focus at $X_i$, are shown. The momentum of these leptons at location $X_i$ increases proportional to the B-field, i.e. leptons at $X_1$ for B=1090G have a momentum $\beta\gamma$ ten times that for leptons at $X_1$ and B=109 G.*

It is seen immediately that the invariance of both, the geometric location of arcs and of the foci on these arcs under variation of B, introduces significant simplifications in the configuration of the 2D-PSD lepton detectors in the detector plane; this apparently entails simplifications in the calibration procedures as well.

For momenta well above $p_1$ the determination of the DDCS is not affected by the characteristics of lepton trajectories as long as the azimuthal twist of the trajectory from origin to the detector plane, as illustrated in fig. 19 and fig. 20, is taken into consideration. This azimuthal twist will only affect the location of the plane of emission but not the determination of the gyro radius $r_g$.

From figs 24 and 9 we note that for large laboratory momenta of leptons below the kinematic maximum only small *transverse* momenta contribute. E.g., for 1087G the largest perpendicular excursion a lepton with emitter frame energy 1MeV will have to the arcs $C^+/C^-$ in the detector plane is 46mm corresponding to near $90^0$ emission in the emitter frame. For locations between the momentum foci on the arcs $C^+/C^-$ one samples for increasing n of the foci more and more  backward emission in the emitter frame with decreasing fraction of the DDCS.

It is interesting to note that due to the kinematics of a fast moving emitter frame large *transverse* momenta can only appear for small laboratory emission angles $\Theta_{lab}$; large angels $\Theta_{lab}$ are related via the kinematics with very small laboratory momenta. Lepton momenta will thus trace out a narrow region left and right of $C_+$ or $C_-$ and only at large laboratory lepton momenta a broad region on either side of $C_+$ and $C_-$ is kinematically possible.

e)  Differential cross sections in emitter frame

When the 2D PSD detector arrays are suitably distributed along the arcs in the detector plane they selectively cover foci and intermediate regions as needed and so permit to determine unconditional double differential cross sections (DDCS) $\frac{d^2\sigma}{dpd\Omega}$ for electron and positron production. When one applies suitable coincidence conditions for detection of electron-positron pairs the triple and partially the full quadruple differential cross sections 3DCS and 4DCS for free-free pair production may then be derived ( e.g. $3DCS(e^+) = \frac{d^3\sigma}{dp_+d\Omega_+dp_-}$ )  .

As electrons and positrons emitted as lepton pairs are not kinematically coupled, i.e. the relation between emitter frame coordinates and laboratory frame coordinates of one lepton of the pair is entirely independent of the emitter and laboratory frame coordinates of the other lepton, the Jacobi matrix for transformation of the cross sections form the emitter into the laboratory frame is diagonal in the leptons. This means that for calculating the essential Jacobian of the Lorentz transformation we may focus on the DDCS.

For the double differential lepton production cross sections the desired emitter frame cross section (primed variables) is derived from the measured laboratory frame cross section by evaluation the Jacobian of the corresponding Lorentz transformation. K. Dedrick reports [132] that Wolfgang Panofsky was the first to give in high energy physics the transformation for a laboratory DDCS $\frac{d^2\sigma}{dEd\Omega}$ to provide the corresponding emitter frame DDCS $\frac{d^2\sigma'}{dE'd\Omega'}$; he showed that evaluation of the Jacobian

$\begin{vmatrix} \frac{\partial E'}{\partial E}\big|_{cos\theta} & \frac{\partial cos\theta'}{\partial E}\big|_{cos\theta} \\ \frac{\partial E'}{\partial cos\theta}\big|_{E} & \frac{\partial cos\theta'}{\partial cos\theta}\big|_{E} \end{vmatrix}$ leads to the equation (see also appendix 2)

$$\frac{d^2\sigma'}{dE'd\Omega'} = \frac{p}{p'}\frac{d^2\sigma}{dEd\Omega} \quad , \tag{12}$$

a relation of astonishing simplicity.

In the current study of pair production in relativistic collision system using a magnetic toroidal spectrometer the momentum space provides the more suitable frame for the differential cross sections. Once more, after some algebra the determinant for the corresponding Jacobian matrix $\begin{vmatrix} \frac{\partial p'}{\partial p}\big|_{cos\theta} & \frac{\partial cos\theta'}{\partial p}\big|_{cos\theta} \\ \frac{\partial p'}{\partial cos\theta}\big|_{p} & \frac{\partial cos\theta'}{\partial cos\theta}\big|_{p} \end{vmatrix}$ yields for the Lorentz transformation the DDCS in the emitter frame as

$$\frac{d^2\sigma'(p',\Omega')}{dp'd\Omega'} = \frac{\gamma}{\gamma'}\frac{(p')^2}{p^2}\frac{d^2\sigma}{dpd\Omega} \quad , \tag{13}$$

where again, as in the derivation of eq. 12, only the well-known relations for the Lorentz transformation of energy and momentum are used. This surprisingly simple relation was first given by Hagedorn [142] and Dedrick [132]. In the appendix we rederive the explicit expressions for all individual matrix elements of both Jacobians used in derivation of eqs. 12 and 13, because these individual terms are considerably more elaborate than the resulting products appearing in eqs 12 and 13. For the 4DCS then one has the product of the Jacobian determinants for electron and positron.

f) Recoil ions and electrons in a magnetic toroidal spectrometer, impact parameter of free-free pair production

The observation that a toroidal magnetic field separates particles of opposite charge transverse to the bend plane of the spectrometer as illustrated above for electrons and positrons, immediately encourages curiosity as to the suitability of this field configuration for coincident detection of electrons and recoil ions produced in a heavy–ion atom collision. It is evident that besides the different mass the emission direction and momentum have to be considered. In the following we use a symmetric very heavy collision system at adiabatic and near relativistic collision velocities to demonstrate the imaging properties for simultaneous mapping of lepton and recoil ions.

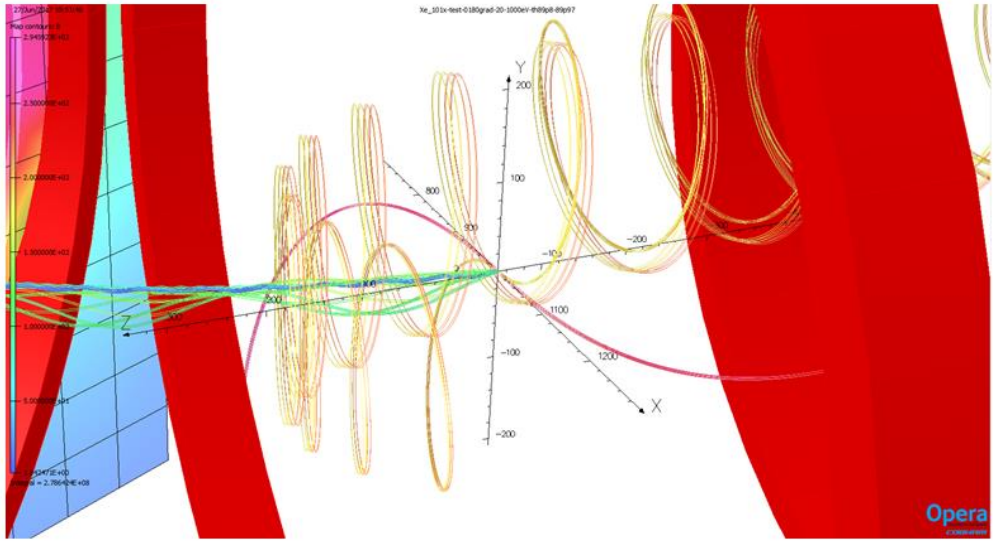

*Fig 26. Simultaneous emission of electrons (blue and green tracks extending from the jet towards the left) and selected recoil ions with emission angels 89.97⁰/89.98⁰ with azimuth 0⁰ and180 energy energies 20 eV (yellow trajectories). Note reflection of recoils in the B=272G field for 20 eV and azimuth 0⁰, but not for azimuth 180⁰. Red trajectories correspond to recoil energies 1000eV with the same emission angles as for 20 eV.*

Theory (28, 29) using coupled channel calculations predicts <b> $\approx$ 100fm, i.e. $E_{kin}$= 0.1AkeV for recoiling target ions ( see also figs. 27 and 28). For B=270 G we find $\rho$= 35 cm for $Au^{40+}$ recoils extracted onto 2D PSDs antiparallel to $v_{proj}$. Bhabha [24, 25] has arrived at a similar result for the relevant impact parameters for lepton pair creation.

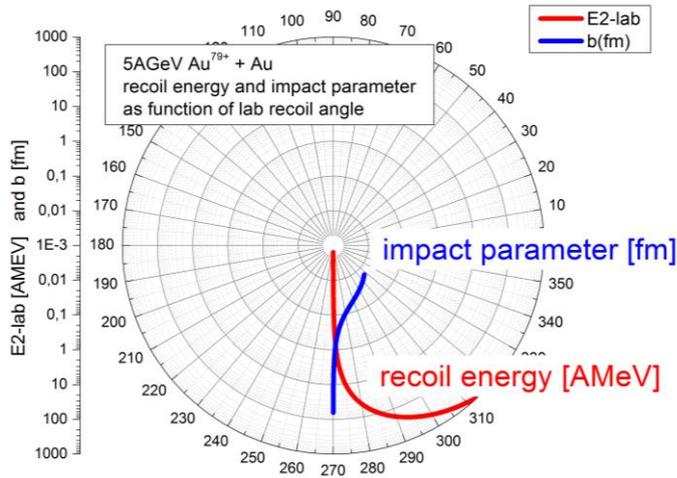

*Fig 27. Polar plot for impact parameter and recoil ion energy as function of laboratory recoil emission angle for 5 AGeV Au79+ + Au. Even for collisions predicted to be dominantly contributing to free-free pair production the recoil ions are emitted with angles very close to 90⁰ with respect to the beam axis.*

More than the high mass, it is the extreme polar emission angle (corresponding to the relevant impact parameters) with respect to the magnetic field direction, which affects the mapping of target recoiling from the collisions. From fig. 26 it is apparent that the position dependent detection of the recoil ions moving in the toroidal field opens the door to extreme sensitive determination of the azimuth of the recoil emission angle. We will analyse the potential for very high resolution spectroscopy employing the extreme azimuthal sensitivity of the spectrometer in a forthcoming paper.

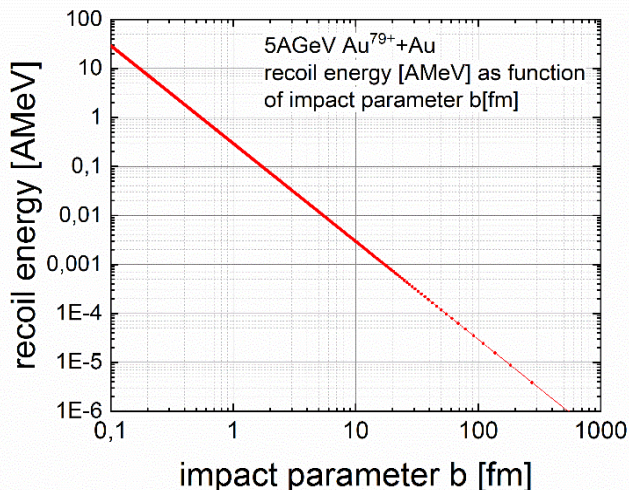

*Fig. 28 Recoil ion energy as function of impact parameter for 5AGeV Au79+ + Au. Note the specific recoil energy of around 50AeV for the predicted most probable impact parameter 100fm facilitating the determination of impact parameters via recoil TOF.*

### 7. Summary

Free-free pair production cross sections in near relativistic ion-atom collisions are predicted to produce lepton pairs over a wide range of momenta and relative emission directions- contrary to photon induced case. Due to considerable computational requirements theory shows only partial DDCS, but no fully 4DCS $\frac{d^4\sigma}{dE_{e^+}d\Omega_{e^+}dE_{e^-}d\Omega_{e^-}}$.

We present an investigation of lepton trajectories in toroidal magnetic fields and we analyze a large acceptance magnetic toroidal sector spectrometer for coincident studies of electrons and positrons from ion-atom collisions .

The toroidal magnetic configuration offers the advantage to image simultaneously the vector momenta of electron and positron of a free-free lepton pair with near $4\pi$ solid angle. This opens the door to selected benchmarks for 3DCS $\frac{d^3\sigma}{dE_{e^+}d\Omega_{e^+}dE_{e^-}}$ and $\frac{d^3\sigma}{dE_{e^-}d\Omega_{e^-}dE_{e^+}}$, at this time there are no experimental 4DCS $\frac{d^4\sigma}{dE_{e^+}d\Omega_{e^+}dE_{e^-}d\Omega_{e^-}}$ for stringent test of theories.

A preliminary explorative analysis of recoil trajectories in the toroidal field indicates that the dynamics of free-free pair production may be accessible via kinematic coincidences of recoiling target ions with lepton pairs.

### Appendix 1: The vector potential in toroidal coordinates

Interest in the behavior of the vector potential **A** inside a toroid did not arise with modern toroidal devices in plasma physics and corresponding Hamiltonian treatment of charged particle dynamics in plasma, but can be traced to Niven's 2$^{nd}$ edition of Maxwell's treatise [131 ]; this first derivation of **A** was, however, shown to be erroneous by Schenkel et al. [100,101].

For a quick visualization concerning the overall configuration of the magnetic vector potential **A** for a current distribution **j,** Maxwell's equations of magnetostatics provide a nice observation, which to our knowledge was first given by Schlomka et al. [ 143 ], but has since been independently reported by other authors [133] as well, because of its great usefulness, e.g. in toroidal configurations. The vector potential **A₁** resulting from a current density distribution **j₁** is in direction and magnitude equal to the magnetic field **B₂** generated by a current density distribution **j₂ ,** when **j₂** equals up to a factor c/4$\pi$ (in Gauß units) the magnetic field **B₁** originating from **j₁**. In other words, the vector potential **A** may be constructed form the field **B** in the same way as **B** is calculated from **j.** For the direction of **A** inside the toroid with a uniform and strictly meridional current distribution they showed that **A** has approximately the same direction as the generating current.

This has been exploited to visualize magnetic vector potentials for various current density distributions deviating from circular shapes [100, 101, 133].

To our knowledge Schenkel et al. [100,101] were then the first to calculate the vector potential **A** inside and outside a toroid using toroidal coordinates, implementing a series expansion of 1/r in toroidal functions given explicitly already in the 19[th] century by Hübschmann [102], for various current density distributions on a toroidal surface. The toroidal (or ring-) functions used in the expansion are the associated Legendre functions with odd-half-integer degree and integer order. With increasing interest in magnetic fields generated by currents in toroidal geometry in plasma physics, eg. of toroidal stellarators, a large number of investigations, besides Schenkel et al. [100],e.g. by Carron et al. [ 133] Mirin [ 141], Haas [ 99a] and Cohn et al. [ 120-122] began to focus on the vector potential inside a toroidal configuration, apparently without knowledge of previous treatises.

Methods for quasi-analytic solutions in terms of a vector potential A were investigated by [141, 99a,135]. Extensive applications of Hamiltonian theory and equations of motion assuring a much broader applicability beyond non-Hamiltonian guiding –center theory has been introduced by [ 95,96,99,106-115,118, 123].

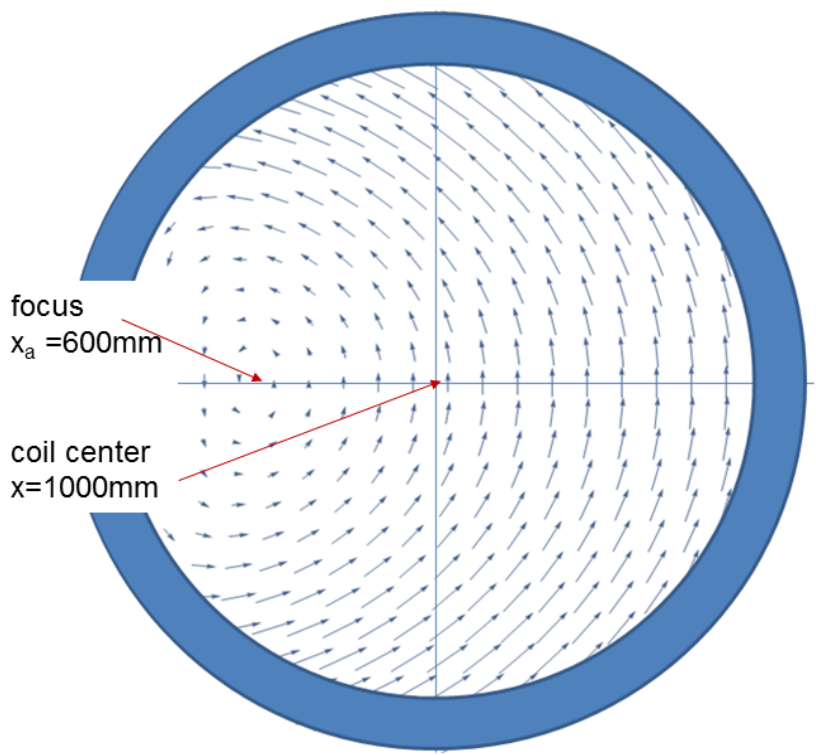

focus
$x_a$ =600mm

coil center
x=1000mm

*Fig. 29 Cut through a toroidal coil, e.g. coil 7 in fig 11, with orientation of the magnetic vector potential **A** inside the toroidal coil configuration with a strictly meridional current distribution on the surface of the coils, calculated using expansions for **A** given by Schenkel [100, 101]. The coil center lies on the toroidal arc as in fig 11; at the focus the toroidal coordinate $\eta$ goes to infinity (see also fig 14a). In the proximity of the inner coil surface the quality of the convergence of the series expansion is decreasing. Lepton trajectories originating at the origin below the paper plane fly towards the observer in the detector plane (see fig. 11); in this figure all detectors are to be imagined to the right of the coil center on the invariant arcs.*

Useful for application in the Hamilton equations, one derives in toroidal coordinates $(\eta, \theta, \psi)$ from the observation of the pure azimuthal B-field from $\vec{B} = \vec{\nabla} \times \vec{A}$ that the magnetic vector potential has only a $\theta$-component $\vec{A} = (A_\eta = 0, A_\theta[\eta, \theta], A_\psi = 0)$ .

Fig 29 gives for the cross section of a toroid (in a geometry as considered in our trajectory calculations with OPERA3D) the vector potential **A** as following from the series expansions of Schenkel et al. [100,101] and Mirin et.[141; we notice a small disagreement with respect to expectations from current density on the inside of the toroidal coil in the proximity to the coil, probably due to slower series convergence.

We have embarked to find suitably parametrized approximations for the numerical vector potential **A** (as displayed in fig 30) to be used in the toroidal Hamiltonian $H = \frac{(\cosh \eta - \cos \theta)^2}{2ma^2}\left[p_\eta^2 + p_\theta^2 + \frac{(p_\psi - A_\psi)^2}{sinh^2 \eta}\right]$ which may be amenable to describe trajectories via the Hamilton equations related to the observed invariants in the lepton trajectories which we found using the numerical techniques in OPERA-3D. Details will be subject of a forthcoming paper.

Besides analysis of particle transport in thermonuclear plasma in tokamaks or other fusion devices also experiments related to the Aharonov-Bohm effect have stimulated a search for expressions for the magnetic vector potential A in toroidal configurations [133].

On the mathematical side it is important to note a few errors in the formulae given in the literature in the context of toroidal coordinates which unfortunately even have propagated into Mathematica.

The defining equations for toroidal functions given in Wolfram World[125] are the erroneous functions from the 9th printing of Abramowitz-Stegun 1st ed [126] which in turn was only later corrected in their 2nd edition; it, was unfortunately propagated into Mathematica 10 and MathWorld. The equations given in the corrected version of Abramowitz- Stegun [127] are in complete agreement with the equations given by Hübschmann[ 102] already in 1890; Mirin [141] has used a slightly different formulation for the toroidal functions.

We also point out, that in Fig, 4.04 in Moon/Spencer [98] illustrating toroidal coordinates the arrow on their θ-coordinate is pointing outward instead of inward. The direction of increasing coordinate theta is displayed correctly in fig. 2.09, which treats the corresponding bipolar coordinates.

**Appendix 2: transformation of cross sections**

For the problem of relating measured energy-angle and/or momentum/angle distributions in the laboratory frame to the corresponding distributions in the emitter frame it is necessary to consider the Jacobians of the Lorentz transformations of the distributions. As the individual terms of the matrix are significantly more elaborate than the final determinant, we give here the explicit expressions for both transformations, for the cross sections differential in the energy and also for those differential in the lepton momentum, as they were first given by Panofsky [132] and e.g. Hagedorn[143].

For evaluating the determinant of the Jacobian
$$\begin{vmatrix} \left.\frac{\partial E'}{\partial E}\right|_{cos\theta} & \left.\frac{\partial cos\theta'}{\partial E}\right|_{cos\theta} \\ \left.\frac{\partial E'}{\partial cos\theta}\right|_{E} & \left.\frac{\partial cos\theta'}{\partial cos\theta}\right|_{E} \end{vmatrix}$$

it is straightforward to derive the four matrix elements

$$J_{11} = \left.\frac{\partial E'}{\partial E}\right|_{cos\theta} = \gamma_0 (1 - \beta_0 \cos\theta \; \frac{E}{p})$$

$$J_{12} = \left.\frac{\partial cos\theta'}{\partial E}\right|_{cos\theta}$$

$$= \frac{\gamma_0}{p'c}(\frac{E\cos\theta}{pc} - \beta_0 - \frac{E'}{p'c}\cos\theta' + \frac{E'E}{p'p}\beta_0\cos\theta\cos\theta')$$

$$J_{21} = \left.\frac{\partial E'}{\partial\cos\theta}\right|_{E} = -\gamma_0\beta_0 pc$$

$$J_{22} = \frac{\partial \cos \theta'}{\partial \cos \theta}\bigg|_E = \gamma_0 \frac{p}{p'}(1 + \frac{E'\beta_0}{p'c}\cos \theta')$$

One finds $J_{11}J_{22} - J_{12}J_{21} = \frac{p}{p'}$. From this directly follows eq.12:

$$\frac{d^2\sigma'}{dE'd\Omega'} = \frac{p}{p'}\frac{d^2\sigma}{dEd\Omega}. \qquad (12)$$

For transformation of the momentum differential cross sections $\frac{d^2\sigma}{dpd\Omega}$ one has to

evaluate elements for the determinant of the corresponding Jacobian

$$\begin{vmatrix} \frac{\partial p'}{\partial p}\big|_{cos\theta} & \frac{\partial cos\theta'}{\partial p}\big|_{cos\theta} \\ \frac{\partial p'}{\partial cos\theta}\big|_{p} & \frac{\partial cos\theta'}{\partial cos\theta}\big|_{p} \end{vmatrix},$$

here one derives the following  matrix elements

$$J_{11}^p = \frac{\partial p'}{\partial p} = \frac{E'}{p'c}\frac{pc}{E}\gamma_0(1 - \beta_0 \cos \theta \frac{E}{pc})$$

$$J_{12}^p = \frac{\partial \cos \theta'}{\partial p} =$$

$$\frac{p}{p'}\frac{\gamma_0}{E}(\frac{E}{pc}\cos \theta - \beta_0 - \frac{E'}{p'c}\cos \theta' + \frac{\beta_0 E'E}{p'cpc}\cos \theta' \cos \theta)$$

$$J_{21}^p = \frac{\partial p'}{\partial \cos \theta} = -\gamma_0 \beta_0 (\frac{pE'}{p'})$$

$$J_{22}^p = \frac{\partial \cos \theta'}{\partial \cos \theta} = \gamma_0 \frac{p}{p'}(1 + \frac{E'\beta_0}{p'c}\cos \theta')$$

Now one finds $\quad J_{11}^p J_{22}^p - J_{12}^p J_{21}^p = \frac{(p')^2}{p^2}\frac{E}{E'} = \frac{(p')^2}{p^2}\frac{\gamma}{\gamma'}.$

From this now expression eq. 13 for obtaining the DDCS in the emitter frame from the DDCS in the laboratory frame

$$\frac{d^2\sigma'(p',\Omega')}{dp'd\Omega'} = \frac{\gamma}{\gamma'}\frac{(p')^2}{p^2}\frac{d^2\sigma}{dpd\Omega} \qquad (13)$$

is directly derived.

For convenience we add the relation for the double differential cross sections in energy and momentum space:

$$\frac{d^2\sigma}{dEd\Omega} = \frac{d^2\sigma}{dpd\Omega}\frac{E}{pc} \tag{14}$$

## Acknowledgments

Within this project Yu. A. Litvinov has received funding from the European Research Council (ERC) under the European Union's Horizon 2020 research and innovation programme (grant agreement No 682841 "ASTRUm")
 P.M. Hillenbrand acknowledges support by DFG project HI 2009/1-1.